# Prevailing climatic trends and runoff response from Hindukush-Karakoram-Himalaya, upper Indus basin

S. Hasson[1,2], J. Böhner[1], V. Lucarini[2,3]

[1] CEN, Centre for Earth System Research and Sustainability, Institute for Geography, University of Hamburg, Hamburg, Germany

[2] CEN, Centre for Earth System Research and Sustainability, Meteorological Institute, University of Hamburg, Hamburg, Germany

[3] Department of Mathematics and Statistics, University of Reading, Reading, UK

Correspondence to: S. Hasson (shabeh.hasson@uni-hamburg.de)

## Abstract

Largely depending on meltwater from the Hindukush-Karakoram-Himalaya, withdrawals from the upper Indus basin (UIB) contribute to half of the surface water availability in Pakistan, indispensable for agricultural production systems, industrial and domestic use and hydropower generation. Despite such importance, a comprehensive assessment of prevailing state of relevant climatic variables determining the water availability is largely missing. Against this background, we present a comprehensive hydro-climatic trend analysis over the UIB, including for the first time observations from high-altitude automated weather stations. We analyze trends in maximum, minimum and mean temperatures (Tx, Tn, and Tavg, respectively), diurnal temperature range (DTR) and precipitation from 18 stations (1250-4500 m asl) for their overlapping period of record (1995-2012), and separately, from six stations of their long term record (1961-2012). We apply Mann-Kendall test on serially independent time series to assess existence of a trend while true slope is estimated using Sen's slope method. Further, we statistically assess the spatial scale (field) significance of local climatic trends within ten identified sub-regions of UIB and analyze whether the spatially significant (field significant) climatic trends qualitatively agree with a trend in discharge out of corresponding sub-region. Over the recent period (1995-2012), we find a well agreed and mostly field significant cooling (warming) during monsoon season i.e. July-October (March-May and November), which is higher in magnitude relative to long term trends (1961-2012). We also find general cooling in Tx and a mixed response in Tavg during the winter season and a year round decrease in DTR, which are in direct contrast to their long term trends. The observed decrease in DTR is stronger and more significant at high altitude stations (above 2200 m asl), and mostly due to higher cooling in Tx than in Tn. Moreover, we find a field significant decrease (increase) in late-monsoonal precipitation for lower (higher) latitudinal regions of Himalayas (Karakoram and Hindukush), whereas an increase in winter precipitation for Hindukush, western- and whole Karakoram, UIB-Central, UIB-West, UIB-West-upper and whole UIB regions. We find a spring warming (field significant in March) and drying (except for Karakoram and its sub-regions), and subsequent rise in early-melt season flows. Such early melt response together with effective cooling during monsoon period subsequently resulted in a substantial drop (weaker increase) in discharge out of higher (lower) latitudinal regions (Himalaya and UIB-West-lower) during late-melt season, particularly during July. These discharge tendencies qualitatively differ to their long term trends for all regions, except for UIB-West-upper, western-Karakorum and Astore. The observed hydroclimatic trends, being driven by certain changes in the monsoonal system and westerly disturbances, indicate dominance (suppression) of nival (glacial) runoff regime, altering substantially the overall hydrology of UIB in future. These findings largely contribute to address the hydroclimatic explanation of the 'Karakoram Anomaly'.

## 1 Introduction

The hydropower generation has key importance in minimizing the on-going energy crisis in Pakistan and meeting country's burgeoning future energy demands. In this regard, seasonal water availability from the upper Indus basin (UIB) that contributes to around half of the annual average surface water availability in Pakistan is indispensable for exploiting 3500 MW of installed hydropower potential at country's largest Tarbela reservoir immediate downstream. This further contributes to the country's agrarian economy by meeting extensive irrigation water demands. The earliest water supply from the UIB after a long dry period (October to March) is obtained from melting of snow (late-May to late-July), the extent of which largely depends upon the accumulated snow amount and concurrent temperatures (Fowler and Archer, 2005; Hasson et al., 2015). Snowmelt runoff is then overlapped by the glacier melt runoff (late-June to late-August), the magnitude of which primarily depends upon the melt season temperatures (Archer, 2003). The snow and glacier melt runoff, originating from the Hindukush-Karakoram-Himalaya (HKH) Ranges, together constitute around 70-80% of the mean annual water available from the UIB (SIHP, 1997; Archer and Fowler 2004; Immerzeel et al., 2009). Contrary to large river basins of the South and Southeast Asia that feature extensive summer monsoonal wet regimes downstream, the lower Indus basin is mostly arid and hyper-arid and much relies upon the meltwater from the UIB (Hasson et al. 2014b).

Climate change is unequivocal and increasingly serious concern due to its recent acceleration. For instance, previous three decades were consecutively warmer at a global scale since 1850, while a period of 1983-2012 in the Northern Hemisphere has been estimated as the warmest since last 1400 years (IPCC, 2013). Such globally averaged warming signal, however, is spatially heterogeneous and not necessarily synchronous among different regions (Yue and Hashino, 2003; Falvey and Garreaud, 2009). Similarly, local impacts of the regionally varying climate change can differ substantially, depending upon the local adaptive capacity, exposure and resilience (Salik et al., 2015), particularly for the sectors of water, food and energy security. In view of high sensitivity of mountainous environments to climate change and the role of meltwater as an important control for UIB runoff dynamics, it is crucial to assess the prevailing climatic state and subsequent water availability from the UIB. Several studies have been performed in this regard. For example, Archer and Fowler (2004) have analyzed trend in precipitation from four stations within the UIB and found a significant increase in winter, summer and annual precipitation during the period 1961-1999. By analyzing the temperature trend for the similar period, Fowler and Archer (2006) have found a significant cooling in summer and a warming in winter, within the UIB. Sheikh et al. (2009) documented a significant cooling of mean temperatures during the monsoon period (July-September), and consistent warming during the pre-monsoonal period (April-May). They have found a significant increase in monsoonal precipitation while non-significant changes for the rest of year. Khattak et al. (2011) have found winter warming, summer cooling (1967-2005), but no definite pattern for precipitation. It is noteworthy that reports from the above mentioned studies are based upon at least a decade old data records. Analyzing updated data for last three decades (1980-2009), Bocchiola and Diolaiuti (2013) have suggested that winter warming and summer cooling trends are less general than previously thought, and can be clearly assessed only for Gilgit and Bunji stations, respectively. They found an increase in precipitation over Chitral-Hindukush and northwest Karakoram regions and decrease in precipitation over the Greater Himalayas within the UIB, though most of precipitation changes are statistically insignificant. By analyzing temperature record only, Río et al. (2013)

also reported dominant warming during March and pre-monsoonal period instead during the winter season, consistent with findings of Sheikh et al. (2009).

The analysis from the above mentioned studies are mostly based upon only a sub-set of half dozen manual, valley-bottom, low-altitude stations being maintained by Pakistan Meteorological Department (PMD - Hasson et al., 2015). Contrary to these low-altitude stations, stations at high altitude in South Asia mostly feature opposite signs of climatic change and extremes, possibly influenced by the local factors (Revadekar et al., 2013). Moreover, bulk of the UIB stream flow is contributed from the active hydrologic altitudinal range (2500-5500 m asl) when thawing temperatures migrate over and above 2500 m asl (SIHP, 1997). In view of large altitudinal dependency of the climate signals, data from low-altitude stations, though extending back into the first half of $20^{th}$ century, are not optimally representative of the hydro-meteorological conditions prevailing over the UIB frozen water resources (SIHP, 1997). Thus, the assessment of the climatic trends over UIB has been much restricted by the limited availability of the high-altitude and most representative observations as well as their accessibility, so far.

Amid above mentioned studies, Archer and Fowler (2004), Fowler and archer (2006) and Sheikh et al. (2009) have used linear least square method for trend analysis. Though such parametric tests more robustly assess the existence of a trend as compared to the non-parametric trend tests (Zhai et al., 2005), they need the sample data to be normally distributed, which is not always the case for the hydro-meteorological observations (Hess et al., 2001; Khattak et al., 2011). In this regard a non-parametric test, such as, Mann Kendall (MK - Mann, 1945; Kendall, 1975) test is a pragmatic choice, which has been extensively adopted for the hydro-climatic trend analysis (Kumar et al., 2009 and 2013). The above mentioned studies of Khattak et al. (2011), Río et al. (2013) and Bocchiola and Diolaiuti (2013) have used the non-parametric MK test in order to confirm the existence of a trend along with Theil-Sen (TS - Theil, 1950; Sen, 1968) slope method to estimate true slope of a trend.

Most of the hydro-climatic time series contain a red noise because of the characteristics of the natural climate variability, and thus, are not serially independent (Zhang et al., 2000; Yue et al., 2002 & 2003; Wang et al., 2008). On the other hand, the MK statistic is highly sensitive to serial dependence of a time series (Yue and Wang, 2002; Yue et al., 2002 & 2003; Khattak et al., 2011). For instance, variance of the MK statistic $S$ increases (decreases) with the magnitude of a significant positive (negative) auto-correlation of the time series, which leads to an overestimation (underestimation) of a trend detection probability (Douglas *et al*., 2000; Yue et al., 2002 and 2003; Wu et al., 2007; Rivard and Vigneault, 2009). To eliminate such affect, von Storch (1995) and Kulkarni and von Storch (1995) proposed a pre-whitening procedure that suggests the removal of a lag-1 auto-correlation prior to applying the MK-test. Río et al. (2013) have analyzed the trends using a pre-whitened (serially independent) time series. This procedure, however, is particularly inefficient when a time series features a trend or it is serially dependent negatively (Rivard and Vigneault, 2009). In fact, presence of a trend can lead to the false detection of a significant positive (negative) auto-correlation in a time series (Rivard and Vigneault, 2009), removing which through the pre-whitening procedure may remove (inflate) the portion of a trend, leading to an underestimation (overestimation) of the trend detection probability and the trend magnitude (Yue and Wang, 2002; Yue et al., 2003). In order to address this problem, Yue et al. (2002) have proposed a modified pre-whitening procedure, which is called trend free pre-whitening (TFPW). In this method, a trend component is separated before the pre-whitening procedure is applied, and after the pre-whitening procedure, the resultant time series is blended together with the pre-identified trend component for further application of the MK-test. Khattak et al. (2011) have applied TFPW procedure to make time series serially independent before trends analysis. The TFPW method takes an advantage of the fact that estimating auto-correlation coefficient from a detrended time series yields its more accurate magnitude for the pre-whitening procedure (Yue et al., 2002). However, prior estimation of a trend may also be influenced by the presence of a serial correlation in a time series in a similar way the presence of a trend contaminates the estimates of an auto-correlation coefficient (Zhang et al., 2000). It is, therefore, desirable to estimate most accurate magnitudes of both trend and auto-correlation coefficient, in order to avoid the influence of one on the other.

The UIB observes contrasting hydro-meteo-cryospheric regimes mainly because of the complex terrain of the HKH ranges and sophisticated interaction of prevailing regional circulations (Hasson et al., 2014a and 2015). The sparse (high and low altitude) meteorological network in such a difficult area neither covers fully its vertical nor its horizontal extents - it may also be highly influenced by the complex terrain features and variability of the meteorological events. Under such scenario, tendencies ascertained from the observations at local sites further need to be assessed for their field significance. This will yield a dominant signal of change and much clear understanding of what impacts the observed conflicting climate change will have on the overall hydrology of the UIB and of its sub-regions. However, similar to the sequentially dependent local time series, the spatial-/cross-correlation amid the station network within a region, possibly due to the influence of a common climatic phenomenon and/or of similar physio-geographical features (Yue and Wang, 2002), anomalously increases the probability of detecting the field significance of local trends (Yue et al., 2003; Lacombe and McCarteny, 2014). Such effect of cross/spatial correlation of a station network should be eliminated while testing the field significance of local trends as proposed by several studies (Douglas et al., 2000; Yue and Wang, 2002; Yue et al., 2003)

In this study, we present a first comprehensive and systematic hydro-climatic trend analysis for the UIB based upon updated dataset from ten stream flow and six low altitude meteorological stations studied earlier, and by including for the first time, observations from 12 high-altitude automatic weather stations from the HKH ranges within the UIB. We apply a widely used non-parametric MK trend test over the serially independent time series, obtained through a pre-whitening procedure, for ensuring the existence of a trend where the true slope of an existing trend is estimated by the Sen's slope method. In pre-whitening, we remove the negative/positive lag-1 autocorrelation that is optimally estimated through an iterative procedure, thus, the pre-whitened time series features the same trend as of the original time series. Here, we investigate the climatic trends on monthly time scale in addition to seasonal and annual time scales, first in order to present more comprehensive picture and secondly to circumvent the loss of intra-seasonal tendencies due to an averaging effect. In view of the contrasting hydrological regimes of UIB due to its complex terrain, highly concentrated cryosphere and the form, magnitude and seasonality of moisture input associated with two distinct modes of prevailing large scale circulation; westerly disturbances and summer monsoon, we decided to investigate in detail the field significance of the local scale climatic trends. In such regards, we divide the whole UIB into ten regions, considering its diverse hydrologic regimes, HKH topographic divides and installed hydrometric station network. Such regions are Astore, Hindukush (Gilgit), western-Karakoram (Hunza), Himalaya, Karakoram, UIB-Central, UIB-West, UIB-West-lower, UIB-West-upper and UIB itself. Provided particular region abodes more than one meteorological station, individual climatic trends within the region were tested for their field significance based upon number of positive/negative significant trends (Yue et al., 2003). Field significant trends are in turn compared qualitatively with the trends of outlet discharge from the corresponding regions in order to furnish physical attribution to statistically identified regional signal of change. Our results, presenting prevailing state of the hydro-climatic trends over the HKH region within the UIB, contribute to the hydroclimatic explanation of the 'Karakoram Anomaly', provide right direction for the impact assessment and modelling studies, and serve as an important knowledge base for the water resource managers and policy makers in the region.

**2 Upper Indus basin and its sub-basins**

The UIB is a unique region featuring a complex HKH terrain, distinct physio-geographical features, conflicting signals of climate change and subsequently contrasting hydrological regimes. The basin extending from the western Tibetan Plateau in the east to eastern Hindu Kush Range in the west, hosts mainly the Karakoram Range in the north and western Himalayan massif (Greater Himalaya) in the south (Fig. 1). It is a transboundary basin, sharing borders with Afghanistan in the west, China in the north and India in the east. The total drainage area of UIB has long been overestimated (e.g. Young and Hewitt, 1988; Alford, 2011; Sharif et al., 2013; Hasson et al., 2014a) - owing to an automated basin delineation procedure based on remotely sensed elevation datasets - featuring a large offset from the original estimates reported by the Surface Water Hydrology Project (SWHP) of the Water and Power Development Authority (WAPDA), Pakistan, that maintains the basin. Here, we have precisely calculated the area of UIB at Besham Qila from the gap-filled 90-meter shuttle radar topographic mission (SRTM) digital elevation model (DEM). For this we have first calculated the basin area using an automated watershed delineation procedure. We have then excluded the adjoining closed-basin areas, for instance, Pangong Tso basin (Khan et al., 2014). Our estimated area of the UIB at Besham Qila is around 163,528 km$^2$, which is, so far, in best agreement with the actual area surveyed and reported by SWHP, WAPDA i.e. 162,393 km2. According to newly delineated basin boundary, UIB is located within the geographical range of 31-37$^o$ E and 72-82$^o$ N, hosting three gigantic massifs, such as, the Karakoram (trans-Himalaya), eastern part of the Hindukush and western part of the Greater Himalaya. A remarkable diversity of the hydro-climatic configurations in UIB is predominantly determined by complex orography of these HKH ranges and the geophysical features, such as presence of frozen water reservoirs. Based on the Randolph Glacier Inventory version 4.0 (RGI4.0 - Pfeffer et al., 2014), these ranges collectively host around 11,000 glaciers, with the Karakoram Range hosting the largest portion. The total area under glaciers and permanent ice cover is around 18,500 km², which is more than 11% of the total surface area of the basin. Around 46 % of the UIB falls within the political boundary of Pakistan, containing around 60 % of the permanent cryospheric extent. The snow coverage within the UIB ranges from 3 to 67% of the total area.

The hydrology of UIB is dominated by the precipitation regime associated with the mid-latitude western disturbances. These western disturbances are the lower-tropospheric extra-tropical cyclones, which are originated and/or reinforced over the Atlantic Ocean or the Mediterranean and Caspian Seas and transported over the UIB by the southern flank of the Atlantic and Mediterranean storm tracks (Hodges et al., 2003; Bengtsson et al., 2006). The western disturbances intermittently transport moisture over the UIB mainly in solid form throughout the year, though their main contribution comes during winter and spring (Wake, 1989; Rees and Collins, 2006; Ali et al., 2009; Hewitt, 2011; Ridley et al., 2013; Hasson et al., 2013 & 2015). Such contributions are anomalously higher during the positive phase of the north Atlantic oscillation (NAO), when southern flank of the western disturbances intensifies over Iran and Afghanistan because of the heat low there, causing additional moisture input to the region from the Arabian Sea (Syed et al., 2006). Similar positive precipitation anomaly is evident during the warm phase of the El Niño–Southern Oscillation (ENSO - Shaman and Tziperman, 2005; Syed et al., 2006). In addition to westerly precipitation, UIB also receives contribution from the summer monsoonal offshoots, which crossing the main barrier of the Greater Himalayas (Wake, 1989; Ali et al., 2009; Hasson et al., 2015), precipitate moisture over higher (lower) altitudes in the solid (liquid) form (Archer and Fowler, 2004). Such occasional incursions of the monsoonal system and the dominating westerly disturbances, largely controlled by the complex HKH terrain, define the contrasting hydro-climatic regimes within the UIB. For the mean annual precipitation, Hasson et al. (2014b) has recently provided a most comprehensive picture of the moisture input to the HKH region within the northern Indus Basin from 36 low-/high-altitude stations, up to an elevation of 4500 m asl. According to their estimates, mean annual precipitation within the UIB ranges from less than 50 mm at Gilgit station to above 1000 mm at Skardu station. Within the Karakoram Range, mean annual precipitation ranges between 200 to 700 mm at Khunjrab and Naltar stations; within the western Himalayas it ranges from 150 to above 1000 mm at Astore and Skardu stations; and within the Hindukush from less than 50 to 400 mm at Gilgit and Ushkore stations, respectively. The glaciological studies however suggest substantially large amount of snow accumulation that account for 1200-1800 mm (Winiger et al., 2005) in Bagrot valley and above 1000 mm over the Batura Glacier (Batura Investigation Group, 1979) within the western Karakoram, and more than 1000 mm and, at few sites above 2000 mm over the Biafo and Hispar glaciers (Wake, 1987) within the central Karakoram.

Within the UIB, the Indus River and its tributaries are gauged at ten key locations, rationally dividing it into various sub-basins namely Astore, Gilgit, Hunza, Shigar and Shyok (Fig. 2). These basins feature distinct hydrological regimes, which are linked with the main source (snow and glacier) of their melt-water generation and can be differentiated by its strong correlation with the climatic variables. For instance, previous studies (Archer 2003; Fowler and Archer, 2006) have separated the snow-fed (glacier-fed) sub-basins of UIB on the basis of their; 1) smaller (larger) glacier coverage and, 2) strong runoff correlation with previous winter precipitation (concurrent temperatures) from low altitude stations. Based on such division, Astore (within the western Himalayan Range) and Gilgit (within the eastern Hindukush Range) basins are considered mainly the snow-fed basins while the Hunza, Shigar and Shyok (within the Karakoram Range) are considered as mainly glacier-fed basins. Since the low-altitude stations do not measure snowfall, such correlation analysis is actually based on winter rainfall, which is not a dominant source of moisture input to the UIB. In fact, unravelling the contrasting hydrological regimes that feature distinct source of melt-water is quite straight forward based on the timing of maximum runoff production (Sharif et al., 2013). Nevertheless, strong influence of the climatic variables on the generated runoff within and from the UIB suggests vulnerability of spatio-temporal water availability to climate change. This is why the UIB discharge features high variability - the maximum mean annual discharge is around an order of magnitude higher than its minimum mean annual discharge, in extreme cases. The mean annual discharge from UIB is around 2400 $m^3s^{-1}$, which contributes to around 45 % of the total surface water availability within Pakistan. The UIB discharge contribution mainly comes from the snow and glacier melt thus concentrates mainly within the melt season (April – September). During the rest of year, melting temperatures remain mostly below the active hydrologic elevation range, resulting in minute melt runoff (Archer, 2004). The characteristics of UIB and its sub-basins are summarized in Table 1. Here, we briefly discuss the sub-basins of UIB.

The Shyok sub-basin located between 33.5-35.7$^o$ E and 75.8-79.8$^o$ N in eastern part of the Karakoram Range constitutes the eastern UIB. The drainage area of Shyok basin has long been overestimated by number of studies, which in fact lead to overestimation of UIB drainage area. This has serious implications for studies, particularly those modelling impacts of climate change on water availability in absolute terms (Immerzeel et al., 2009). According to our updated estimates, which are in best agreement with the SWHP, WAPDA, its drainage area is around 33,000 km². Based on such drainage estimate, the basin elevation range, derived from gap-filled 90 meter SRTM DEM, is 2389-7673 m asl. Based on RGI4.0 (Pfeffer et al., 2014), approximately 24% of the basin area is under the glacier and permanent ice cover, hosting around 42 % of the total glacier cover within the UIB. Westerly disturbances are mainly responsible for moisture input to the Shyok basin; however one third of the solid moisture input comes from the summer monsoon system (Wake, 1989). Mean annual precipitation from the only available high-altitude station Hushe is around 500 mm. The mean annual discharge contribution of 360 m³s⁻¹ is mainly constituted from the snow and glacier melt, which contributes around 15 % to the UIB discharge.

The Shigar sub-basin lies within the central Karakoram Range, coordinated between 74.8-76.8° E and 35.2-36.2° N. Its elevation range is 2189-8448 m asl. Around one third of the basin area lies above 5000 m asl. The basin area is around 7000 km², of which around one third is covered by glaciers, including some of those among the largest in the world. The basin receives its main moisture from the westerly disturbances during the winter and spring season in solid form, however, occasional summer monsoonal incursions drop moisture to the upper reaches and influence the overall hydroclimatology of the basin. The mean annual precipitation input ranges between 450 mm at Shigar high-altitude station to above 1000 mm at nearby low-altitude Skardu station. Representing only the basin below 2400 m asl, these precipitation amounts are quite small compared to those reported by the glaciological studies. The snow cover ranges between 25±8 and 90±3% (Hasson et al., 2014b). The discharge from the Shigar basin mainly comprises of slow runoff (snow and glacier melt runoff) and is estimated to be around 200 m³s⁻¹, which is around 9 % of the mean annual discharge at UIB Besham Qila.

The Gilgit sub-basin (between 35.8-37° E and 72.5-74.4° N) encompasses eastern part of the Hindukush Range and drains southeastward into the Indus River. Gilgit River is measured at Gilgit hydrometric station, right after which the Hunza River confluence with the Gilgit River at Alam Bridge. The drainage area of the basin corresponds to more than 12000 km² with an elevation range of 1481-7134 m asl. Around 7 % of the basin area is under glacier and permanent ice cover, accounting for 4% of the UIB cryospheric extent. The Gilgit basin receives its precipitation from both westerly disturbances and summer monsoon system, which amounts less than 50 mm at Gilgit station to more than 350 mm at Ushkore station (Hasson et al., 2014b). Snow cover in the basin ranges between 3±1 and 90±4% (Hasson et al., 2014b). Discharge mainly depends upon the snowmelt, followed by the glacier melt and rainfall. Mean annual discharge out of Gilgit basin is around 300 m$^3$s$^{-1}$, which contributes around 12% to the UIB mean annual discharge.

The Hunza sub-basin abodes mainly the western part of the Karakoram Range and covers an area of 13734 km$^2$. It also includes area of east and southeastward draining Hindukush massifs. It is located within the coordinates 35.9-37.2$^o$ E and 74-75.8$^o$ N. The elevation range of basin is 1420-7809 m asl where one third of the basin lies above 5000 m asl, alike Shigar basin. Around 28 % of its total surface area is covered by glacier and permanent ice (Pfeffer et al., 2014), which is almost 21% of the permanent cryospheric extent of UIB. Mean snow cover ranges from *17±6* to 83±4 % of the total basin area during the period 2001-2012 (Hasson et al., 2014b). Mean annual moisture input ranges from 200 at Khunjrab station to 700 mm at Naltar station during the period 1995-2012 (Hasson et al., 2014b). The mean annual discharge for the period 1966-2010 is 330 m$^3$s$^{-1}$, which contributes approximately 14% to the mean annual discharge of UIB at Besham Qila.

The Astore sub-basin, lying within the southern foothills of western Himalayan extremity, is the only north-facing gauged basin within the UIB, located between 34.7-35.6$^o$ E and 74.3-75.3$^o$ N. It has a drainage area of around 3900 km$^2$ with an elevation range of 1504-8069 m asl, where only a small area lies above 5000 m asl. Almost 14% of the total basin area is covered by permanent ice and glaciers, aboding only 3% of the total within the UIB. Snow cover within the basin ranges from 2±1 to 98±1% (Hasson et al., 2014b). The hydrology of Astore basin is mainly influenced by the westerly solid moisture input, however the basin receives one third of its annual precipitation under the summer monsoon system (Farhan et al., 2014). Mean annual precipitation within the Astore basin ranges from around 140 mm at the rainfall-only low-altitude Astore station to above 800 mm at high altitude Ramma station (Hasson et al., 2014b). The mean annual runoff from Astore basin measured at Dainyor site is around 140 m$^3$s$^{-1}$, which contributes around 6% of the mean annual discharge at UIB Besham Qila.

**3 Data**

### 3.1 Meteorological data

The network of meteorological stations within the UIB is very sparse and mainly limited to within Pakistan's political boundary, where around 20 meteorological stations are being operated by three different data collection organizations. The first network, being operated by PMD, consists of six manual valley based stations that provide only long-term data series, generally starting from first half of the 20th century. However, data before 1960 are scarce and feature large data gaps (Sheikh et al., 2009). Such dataset covers a north-south extent of around 100 km from Gupis to Astore station and east-west extent of around 200 km from Skardu to Gupis station. The altitudinal range of these stations is limited to 1200-2200 m asl only and merely within the western Himalaya and Hindu Kush ranges. Whereas most of the ice reserves of the Indus Basin lie within the Karakoram range (Hewitt, 2011) and above 2200 m asl (Fig. 1). In view of the fact that bulk contribution to the UIB stream flow occurs from the active hydrologic altitudinal range of 2500-5500 m asl when thawing temperatures migrate above 2500 m asl (SIHP, 1997), the low altitude stations are not optimally representative of the hydro-meteorological conditions prevailing over the UIB cryosphere. The EvK2-CNR has installed two meteorological stations in the central Karakoram at higher elevations, which however, provide time series only since 2005. Moreover, the precipitation gauges within PMD and EvK2CNR networks measure only liquid precipitation, while hydrology of the region is dominated by solid moisture input. The third meteorological network within the UIB consists of 12 high altitude automatic weather stations, called Data Collection Platforms (DCPs), which are being maintained by the Snow and Ice Hydrology Project (SIHP) of WAPDA. Data is being observed at hourly intervals and is transferred on real time basis through a Meteor-Burst communication system to the central SIHP office in Lahore. The data is subject to missing values due to rare technical problems, such as 'sensor not working' and/or 'data not received from broadcasting system'. Featuring higher altitude range of 1479-4440 m asl, these DCP stations provide medium-length time series of meteorological observations since 1994/95. Contrary to lower altitude stations, precipitation gauges at DCPs measure both liquid and solid precipitation in mm water equivalent (Hasson et al., 2014b). Moreover, DCPs cover relatively larger spatial extent, such as, north-south extent of 200 km from Deosai to Khunjrab station and east-west extent of around 350 km from Hushe to Shendure stations. Thus, spreading well across the HKH ranges and covering most of the vertical extent of UIB frozen water resources and the active hydrologic altitudinal range, DCPs seem to be well representative of the prevailing hydro-meteorological conditions over the UIB cryosphere, so far. We have collected the daily data for the temperature maximum, temperature minimum (Tx and Tn, respectively) and precipitation of 12 DCP stations for the period 1995-2012 from SIHP, WAPDA. We have also collected the updated record of six low altitude stations from PMD for same set of variables within the period 1961-2012. Details of the collected meteorological observations are listed in Table 2.

**3.2 Discharge data**

The discharge data, being highly sensitive to variations in precipitation, evaporation, basin storage and prevailing thermal regime, describes the overall hydrology and the integrated signal of hydrologic change for a particular watershed. In order to provide physical attribution to our statistically based field significant trend analysis, we have collected the discharge data from SWHP, WAPDA. The project maintains a network of hydrometric stations within the Pakistan region. The upper Indus river flows are being measured first at Kharmong site where the Indus river enters into Pakistan Territory and then at various locations until it enters into the Tarbela Reservoir. The river inflows measuring stations at Tarbela reservoir, and few kilometers above it, at the Besham Qila are usually considered to separate the upper part of the Indus (i.e. UIB) from the rest of basin. The hydrometric station network rationally apportions UIB into smaller units based upon distinct hydrological regimes and magnitude of runoff contributions. Almost five sub-basins are being gauged, from which Shigar gauge is not operational after 2001. Since we take the UIB extent up to the Besham Qila site, we have collected full length of discharge data up to 2012 for all ten hydrometric stations within the UIB. Details of the collected discharge data are given in Table 3 in downstream order. It is pertinent to mention here that discharge data from central and eastern parts of the UIB are hardly influenced by the anthropogenic perturbations. The western UIB is relatively populous and stream flow is used for solo-seasoned crops and domestic use, however, the overall contribution to such use is still negligible (Khattak et al., 2011).

**4 Methods**

Inhomogeneity in climate time series is due to variations in the record that can be ascribed to purely non-climatic factors (Conrad and Pollak, 1950), such as, changes in the station site, station exposure, observational method, and measuring instrument (Heino, 1994; Peterson et al., 1998). Archer and Fowler (2004) and Fowler and Archer (2005 and 2006) have documented that PMD and WAPDA follow standard meteorological measurement practice established in 1891 by the Indian Meteorological Department. Using double mass curve approach, they have found inhomogeneity in the winter minimum temperature around 1977 only at Bunji station among four low altitude stations analyzed. Since climatic patterns are highly influenced by orographic variations and local events within the study region of complex terrain, double mass curve techniques may yield limited skill. Forsythe et al. (2014) have reported the homogeneity of Gilgit, Skardu and Astore stations for annual mean temperature during the period 1961-1990 while Río et al. (2013) have reported the homogeneity for the temperature record from the Gilgit, Gupis, Chillas, Astore and Skardu stations during 1952-2009. Some studies (Khattak et al., 2011; Bocchiola and Diolaiuti, 2013) do not report the quality control or homogeneity of the data used for their analysis.

We have first investigated the internal consistency of the data by closely following Klein Tank et al. (2009) such as situations of below zero precipitation and when maximum temperature was lower than minimum temperature, which found in few were then corrected. Afterwards, we have performed homogeneity test using a standardized toolkit RH-TestV3 (Wang and Feng, 2009) that uses a penalized maximal F-test (Wang et al., 2008) to identify any number of change points in a time series. As no station has yet been reported homogenous at monthly time scale for all variables, and that stations observe large Euclidean distance in a highly complex terrain, we were restricted to perform only a relative homogeneity test, without using a reference time series. We have tested the homogeneity for the monthly mean maximum and minimum temperatures and monthly total precipitation by adopting a most conservative threshold level of 99% for statistical significance. We have found mostly one inhomogeneity in only Tn for the low altitude PMD stations during the period of record, except for the Skardu station (Table 2). Within the 1995-2012 period, such homogeneity in Tn is only valid for Gilgit and Gupis stations. On the other hand, data from DCP stations were found of high quality and homogenous. Only Naltar station has experienced inhomogeneity in Tn during September 2010, which was most probably caused by heavy precipitation event resulted in a mega flood in Pakistan (Houze et al., 2011; Ahmad et al., 2012; Hasson et al., 2013) followed by similar events during 2011 and 2012. Since the history files were not available, we were not sure that any statistically found inhomogeneity in only Tn is real. Therefore, we did not apply any correction to the data and caution the careful interpretation of results based on such time series.

## 4.1 Hydroclimatic trend analysis

We have analyzed trend in the minimum, maximum and mean temperatures (Tn, Tx and Tavg, respectively), diurnal temperature range (DTR – Tx - Tn), precipitation and discharge on monthly to annual time scales. For this, we used a widely applied nonparametric MK statistical test (Mann, 1945; Kendall, 1975) to assess the existence of a trend along with Theil-Sen (TS - Theil, 1950; Sen, 1968) slope method to estimate true slope of an existing trend. For sake of intercomparison between low and high altitude stations, we mainly analyze overlapping length of record from the two datasets (i.e. 1995-2012). However, we additionally analyze full length of record from low altitude stations.

**Mann-Kendall test**

The MK is a ranked based method that tests the significance of an existing trend irrespective of the type of the sample data distribution and whether such trend is linear or not (Yue et al., 2002; Wu et al., 2007; Tabari, H., and Talaee, 2011). Such test is also insensitive to the data outliers and missing values (Khattak et al., 2011; Bocchiola and Diolaiuti, 2013) and less sensitive to the breaks caused by inhomogeneous time series (Jaagus, 2006). The null hypothesis of the MK test states that a sample data $\{X_i, i = 1,2,3 \dots n\}$ is independent and identically distributed, while the alternative hypothesis suggests the existence of a monotonic trend. The MK statistics $S$ are estimated as follows:

$$S = \sum_{i=1}^{n-1} \sum_{j=i+1}^{n} sgn(X_j - X_i) \qquad (1)$$

Where $X_j$ denotes the sequential data, n denotes the data length, and

$$sgn(\theta) = \begin{cases} 1 & if\ \theta > 0 \\ 0 & if\ \theta = 0 \\ -1 & if\ \theta < 0 \end{cases} \qquad (2)$$

provided n ≥ 10, $S$ statistics are approximately normally distributed with the mean, $E$, and variance, $V$, (Mann, 1945; Kendall, 1975) as follows:

$$E(S) = 0 \qquad (3)$$

$$V(S) = \frac{n(n-1)(2n+5) - \sum_{m=1}^{n} t_m m(m-1)(2m+5)}{18} \qquad (4)$$

Here, $t_m$ denotes the number of ties of extent *m*, where tie refers to $X_j = X_i$. The standardized MK statistics, $Z_s$, can be computed as follows:

$$Z_s = \begin{cases} \frac{S-1}{\sqrt{V(S)}} & S > 0 \\ 0 & S = 0 \\ \frac{S+1}{\sqrt{V(S)}} & S < 0 \end{cases} \qquad (5)$$

The null hypothesis of no trend is rejected at a specified significance level, $\alpha$, if $|Z_s| \geq Z_{a/2}$, where $Z_{\alpha/2}$ refers to a critical value of standard normal distribution with a probability of exceedance $\alpha/2$. The positive sign of $Z$ shows an increasing while its negative sign shows a decreasing trend. We have reported the statistical significance of identified trends at 10, 5 and 1% levels by taking $\alpha$ as 0.1, 0.05 and 0.01, respectively.

**Theil-Sen's slope estimation**

Provided the time series features a trend, such trend can be approximated by a linear regression as

$$Y_t = a + \beta t + \gamma_t \qquad (6)$$

Where $a$ is the intercept, $\beta$ is a slope and $\gamma_t$ is a noise process. Such estimates of $\beta$ obtained through a least square method are prone to gross errors and the respective confidence intervals are sensitive to the type of parent distribution (Sen, 1968). We, therefore, have used the Theil–Sen approach (TS - Theil, 1950; Sen, 1968) for estimating the true slope of the existing trend as follows

$$\beta = Median\left(\frac{X_j - X_i}{j - i}\right), \forall\, i < j \qquad (7)$$

The magnitude of $\beta$ refers to a mean change in a considered variable over the investigated time period, while a positive (negative) sign implies an increasing (decreasing) trend.

**Trend-perceptive pre-whitening (TPPW)**

In order to pre-whiten the time series for serial dependence, we have used an approach of von Storch (1995) as modified by Zhang et al (2000). In this approach, one iteratively computes the trend and lag-1 auto-correlation of a time series until the solution converges to most accurate estimates of a trend magnitude and autocorrelation - an absolute difference between the estimates from two consecutive iterations becomes negligible. This approach assumes that the trend ($T_t$) in Eqn. 8 can be approximated as linear ($T_t = \beta.t$). Moreover, one assumes that the noise, $\gamma_t$, can be represented as a $p$th order auto-regressive process, AR($p$) of the signal itself, plus the white noise, $\varepsilon_t$.

$$Y_t = a + T_t + \gamma_t \qquad (8)$$

Since the partial auto-correlations for lags larger than one are generally found insignificant (Zhang et al., 2000; Wang and Swail, 2001), considering only lag-1 auto-regressive processes, $r$, yields Eqn. 8 into:

$$Y_t = a + \beta t + rY_{t-1} + \varepsilon_t \qquad (9)$$

The iterative pre-whitening procedure consists of following steps:

1. In the first iteration, estimate of lag-1 autocorrelation, $r_1$ is computed on the original time series, $Y_t$.
2. Using $r_1$ as $(Y_t - r.Y_{t-1})/(1-r)$, an intermediately pre-whitened time series, $\acute{Y}_t$, is obtained on which first estimate of a trend, $\beta_1$ along with its significance is computed using TS (Theil, 1950; Sen, 1968) and MK (Mann, 1945; Kendall, 1975) methods.
3. The original time series, $Y_t$, is detrended using $\beta_1$ as $(\widehat{Y}_t = Y_t - \beta_1 t)$.
4. In the second iteration, more accurate estimate of lag-1 autocorrelation $r_2$ is estimated on a detrended time series, $\widehat{Y}_t$, obtained in a previous iteration.
5. The original time series $Y_t$, is again intermediately pre-whitened and $\acute{Y}_t$ is obtained.
6. The trend estimate $\beta_2$ is then computed on $\acute{Y}_t$ and the original time series, $Y_t$ is detrended again, yielding $\widehat{Y}_t$.

The procedure has to be reiterated until $r$ is no longer significantly different from zero or the absolute difference between the estimates of $r, \beta$ obtained from the two consecutive iterations becomes less than one percent. If any of the condition is met, let's suppose at the iteration $n$, estimates from the previous iteration (i.e. $r = r_{n-1}, \beta = \beta_{n-1}$) are taken as final. Using these final estimates, Eqn. 10 yields a final pre-whitened time series, $Y_t^w$, which is serially independent and features a same trend as of the original time series, $Y_t$ (Zhang et al., 2000; Wang and Swail, 2001). Finally, the MK-test is applied over the pre-whitened time series, $Y_t^w$ to identify existence of a trend.

$$Y_t^w = \frac{(Y_t - r.Y_{t-1})}{(1-r)} = \hat{a} + \beta t + \epsilon_t, \text{ where } \hat{a} = a + \frac{r.\beta}{(1-r)}, \text{ and } \epsilon_t = \frac{\varepsilon_t}{(1-r)} \qquad (10)$$

**4.2 Field significance of local trends**

The field significance indicates that whether the stations within a particular region collectively exhibit a regional significant trend irrespective of either their individual trends were significant or not (Vogel and Kroll, 1989; Lacombe et al., 2013). For assessing the field significance of local trends, we have divided the whole UIB into further smaller units/regions based on: 1) distinct hydrological regimes identified within the UIB; 2) available installed stream flow network, and; 3) hosted mountain massifs. We have considered the whole Karakoram Range as an area within the natural boundaries of the Hunza, Shigar and Shyok basins, which we then considered as western, central and eastern Karakoram regions, respectively (Fig. 2). Similarly, we have considered the basin area up to Indus at Kharmong as UIB-East, area of Shigar and Shyok basins jointly as UIB-Central, and rest of the UIB area as UIB-West (Fig. 2). We have further divided the UIB-West region into its upper and lower parts, keeping in view relatively large number of stations and distinct hydrological regimes, which have been identified, based on timings of their maximum runoff production, by comparing median hydrographs from each steam flow gauging station. According to such division UIB-West-lower and Gilgit are mainly snow-fed basins while Hunza is mainly glacier-fed basin (Fig. 3). Since most of the Gilgit basin area lies at Hindukush massifs, we call it Hindukush region. Additionally, combined area of lower part of UIB-West and UIB-east is mainly the northward slope of the Greater Himalaya, so we call this combined region as Himalaya. Thus, apart from the gauged basins of Astore, Gilgit, Hunza, Shigar and Shyok, Indus at Kharmong (UIB-East), and UIB itself, we have obtained the regions of Karakoram, Himalaya, UIB-Central, UIB-West, UIB-West-lower and UIB-West-upper, for which discharge was derived from installed gauges.

As mentioned earlier, Shigar discharge time series was limited to 1985-2001 period since afterwards the gauge went non-operational. In order to analyze discharge trend from such an important region, Mukhopadhyay et al. (2014) have first correlated the Shigar discharge with discharge from its immediate downstream Kachura gauge for the overlapping period of record (1985-1998). Then, they applied the estimated monthly correlation coefficients to the post-1998 discharge at Indus at Kachura. This particular method can yield the estimated Shigar discharge, of course assuming that the applied coefficients remain valid after the year 1998. However, in view of the large surface area of more than 113,000 km$^2$ for Indus at Kachura and substantial changes expected in the hydroclimatic trends upstream Shigar gauge, discharge estimated by Mukhopadhyay et al. (2014) merely seems to be a constant fraction of the Kachura discharge, rather than the derived Shigar discharge. On the other hand, we have derived the Shigar discharge by excluding the mean discharge rates of all gauges upstream Shigar gauge, which do not represent the Shigar basin, from its immediate downstream Kachura gauge. Such subtraction of all upstream gauges from immediate downstream gauge was performed for each time step of every time scale analyzed during the period of discharge estimation. Similar methodology has been adopted to derive discharge out of identified ungauged regions, based upon the installed stream flow gauges (Eqn. 11-13, Table 1). In this procedure, however, we assume that regions far from each other (UIB-east and UIB-West-lower) have negligible routing time delay at a mean monthly time scale - our shortest time scale analyzed - and that such an approximation does not further influence the ascertained trends. In other words, we derived the discharge for considered ungauged regions by assuming them in place, since our focus was to assess changes in the discharge contribution out of such regions rather than their influence on the UIB outlet discharge at certain time.

$$Q_{(Central-UIB)} = Q_{(Indus\ at\ Kachura)} - Q_{(Indus\ at\ Kharmong)} \qquad (11)$$

$$Q_{(Western-UIB-L)} = Q_{(UIB)} - Q_{(Indus\ at\ Kachura)} - Q_{(Gilgit\ at\ Alam\ Bridge)} \qquad (12)$$

$$Q_{(Western-UIB)} = Q_{(UIB)} - Q_{(Indus\ at\ Kachura)} \qquad (13)$$

We have analysed field significance for those regions that contain at least two or more stations. In order to eliminate effect of cross/spatial correlation of a station network on assessing the field significance of a particular region, Douglas et al. (2000) have proposed a bootstrap method. This method preserves the spatial correlation within a station network but eliminates its influence on testing the field significance of a trend based on the MK statistic $S$. Similarly, Yue and Wang (2002) have proposed a regional average MK test in which they altered the variance of MK statistic by serial and cross correlations. Lately, Yue et al. (2003) proposed a variant of method proposed by Douglas et al. (2000), in which they considered counts of the significant positive and negative trends - instead of the MK statistic $S$ - as representative variables for testing the field significance of both positive and negative trends separately. This method favourably provides a measure of dominant field significant trend when local positive or negative significant trends are equal in number. Therefore, we have employed the method of Yue et al. (2003) for assessing the field significance. We have used a bootstrap approach (Efron, 1979) to resample the original network 1000 times in a way that the spatial correlation structure was preserved as described by Yue et al. (2003). We have counted both the number of local significant positive and number of significant negative trends, separately for each resampled network dataset using Eqn. 14:

$$C_f = \sum_{i=1}^{n} C_i \tag{14}$$

Where *n* denotes total number of stations within a region and $C_i$ denotes a count for statistically significant trend (at 10% level) at station, *i*. Then, we have obtained the empirical cumulative distributions $C_f$ for both counts of significant positive and counts of significant negative trends, by ranking their corresponding 1000 values in an ascending order using Eqn.15:

$$P(C_f \leq C_f^r) = \frac{r}{N+1} \tag{15}$$

Where *r* is the rank of $C_f^r$ and *N* denotes the total number of resampled network datasets. We have estimated probability of the number of significant positive (negative) trends in actual network by comparing the number with $C_f$ for counts of significant positive (negative) trends obtained from resampled networks (Eqn. 16).

$$P_{obs} = P(C_{f,obs} \leq C_f^r), \text{ where } P_f = \begin{cases} P_{obs} & for\ P_{obs} \leq 0.5 \\ 1 - P_{obs} & for\ P_{obs} > 0.5 \end{cases} \tag{16}$$

At the significance level of 10 %, if expression $P_f \leq 0.1$ is satisfied the trend over a region is considered as field significant.

In addition to investigating statistically the field significance of tendencies in meteorological variables, we have provided physically-based evidence from the stream flow record. We have ascertained the trends in stream flow data (from installed and derived gauges) and compared them with the field significant climatic signal, particularly the temperature trend from the corresponding regions. The qualitative agreement between the two can serve better in understanding the ongoing state of climate over the UIB. Since the most downstream gauge of UIB at Besham Qila integrates the variability of all upstream gauges, it represents the dominant signal of change. Thus, an assessment of statistically based field significance was not required for the stream flow dataset.

We also assess the dependency of local hydroclimatic trends on their latitudinal, longitudinal and altitudinal distribution. Here we mention that we have intentionally avoided the interpolation of data and results in view of limitations of the interpolation techniques in a complex terrain of HKH region (Palazzi et al., 2013; Hasson et al., 2015). Large offset of glaciological reports from the station based estimates of precipitation (Hasson et al., 2014b) further suggests that hydro-climatic patterns are highly variable in space and that the interpolation of data will further add to uncertainty, resulting in misleading conclusions.

## 5 Results

First, we present the results of our analysis based upon a common length of record (i.e. 1995-2012) from PMD and DCP stations (Table 4 and 5, and for the select time scales, in Fig. 4). Then, we compare the trends at low altitude stations over the period 1995-2012 with their long-term trends (1961-2012), in order to investigate any recent development of rate or sign of change in the climatic trends. Here we remind that, we call mainly six PMD stations (1200-2200 m asl) as low altitude stations and all the trends estimated over full-length record as long-term trends (Table 6). Similarly, we call DCP stations from SIHP, WAPDA as high altitude stations (2200-4500 m asl). Within the 1995-2012 period, we also compare the results from low altitude stations against the findings from high altitude stations, in order to present their consistencies and variations. We show in Table 7, the field significant trends in climatic variables and trends in discharge from the corresponding regions.

### 5.1 Hydroclimatic trends

**Mean maximum temperature**

For Tx, we find that certain set of months exhibit a common response of cooling and warming within the annual course of time. Set of these months interestingly are different than those typically considered for seasons, such as, DJF, MAM, JJA, SON for winter, spring, summer and autumn, respectively (Fowler and Archer, 2005 and 2006, Khattak et al, 2011; Bocchiola and Diolaiuti, 2013). For the months of December, January, February and April, stations show a mixed response of cooling and warming tendencies by roughly equal numbers where cooling trend for Rattu in January, for Shendure in February and for Ramma in April are statistically significant (Table 4 and Fig. 4). Though no warming trend has been found to be statistically significant, all low altitude stations, except Gupis, exhibit a warming trend in the month of January. During months of March, May and November, most of the stations exhibit a warming trend, which is statistically significant at five stations (Gilgit, Yasin, Astore, Chillas and Gupis) and relatively higher in magnitude during March. Interestingly, warming tendencies during March are relatively higher in magnitude at low altitude stations as compared to high altitude stations. Most of the stations feature cooling tendencies during July-October (mainly the monsoon period). During such period, we find a statistically significant cooling at five stations (Dainyor, Shendure, Chillas, Gilgit and Skardu) in July, at two stations (Shendure and Gilgit) in August and at twelve stations (Hushe, Naltar, Ramma, Shendure, Ushkore, Yasin, Ziarat, Astore, Bunji, Chillas, Gilgit and Skardu) in September, while there is no significant cooling tendency in October (Table 4 and Fig. 4). Such cooling is almost similar in magnitude from low and high altitude stations and dominates during month of September followed by July because of higher magnitude and statistical significance agreed among large number of stations. Overall, we note that cooling trends dominate over the warming trends. On a typical seasonal scale, insignificant but intra-station agreed cooling in February is averaged out for the winter season, which then generally show a mixed behavior (cooling/warming) where only two stations (Dainyor and Rattu) show a significant cooling. For spring season, there is a high agreement for warming tendencies among the stations, which are significant only at Astore station. Again such warming tendencies during spring are relatively higher in magnitude than those at higher altitude stations. For summer and autumn seasons, most of the stations feature cooling tendencies, which are significant for three stations (Ramma, Shendure and Shigar) in summer and for two stations (Gilgit and Skardu) in autumn. On annual time scale, high altitude stations within Astore basin (Ramma and Rattu) feature significant cooling trend.

While looking only at long term trends (Table 6), we note that summer cooling (warming outside summer) in Tx is less (more) prominent and insignificant (significant) at stations of relatively high (low) elevation, such as, Skardu, Gupis, Gilgit and Astore (Bunji and Chillas). The absence of a strong long-term winter warming contrasts with what found for the shorter period 1995-2012. In fact, strong warming is restricted to spring season mainly during March and May months. Similarly, long-term summer cooling period of June-October has been shortened to July-October.

**Mean minimum temperature**

The dominant feature of Tn is the robust winter warming in Tn during November-June, which is found for most of the stations (Table 4 and Fig. 4). Contrary to warming in Tx, warming trend in Tn is higher in magnitude among the high altitude stations than among the low altitude stations. During the period of July-October, we found a significant cooling of Tn at four stations (Gilgit, Naltar, Shendure and Ziarat) in July, at eight stations (Hushe, Naltar, Ushkore, Yasin, Ziarat, Astore, Chillas and Gilgit) in September and only at Skardu in October. In August, stations show warming tendencies, which are relatively small in magnitude and only significant at Gilgit station. Similar to Tx, cooling in Tn during July-October dominates during the month of September suggesting a relatively higher magnitude and larger number of significant trends (Fig. 4). Also, such cooling features more or less similar magnitude of a trend among high and low altitude stations as for Tx. Similarly, cooling trends in Tn mostly dominate over the warming trends as in case of Tx. On a typical seasonal scale, winter and spring seasons feature warming trends, while summer season exhibit cooling trend and there is a mixed response for the autumn season. Warming trend dominates during the spring season. Here, we emphasize that a clear signal of significant cooling in September has been lost while averaging it into October and November months for autumn season. This is further notable from the annual time scale, on which a warming trend is generally dominated that is statistically significant at five stations (Deosai, Khunjrab, Yasin, Ziarat and Gilgit). The only significant cooling trend on annual time scale is observed at Skardu station.

While looking only at low altitude stations (Table 6), we note that long term non-summer warming (summer cooling) in Tn is less (more) prominent and insignificant (significant) at stations of relatively high (low) elevation, such as, Skardu, Gupis, Gilgit and Astore (Bunji and Chillas).

**Mean temperature**

Trends in Tavg are dominated by trends in Tx during July-October while these are dominated by Tn, during the rest of year (Table 4-5). Similar to Tx, the Tavg features a significant cooling in July at four stations (Dainyor, Naltar, Chillas and Skardu), in September at ten stations (Hushe, Naltar, Rama, Shendure, Ushkore, Yasin, Ziarat, Astore, Chillas and Skardu) and in October only at Skardu station (Table 5 and Fig. 4). In contrast, we have observed a significant warming at Ziarat station in February, at five stations (Deosai, Dainyor, Yasin, Astore and Gupis) in March and at three stations (Khunjrab, Gilgit and Skardu) in November. However, trend analysis on typical seasonal averages suggest warming of winter and spring seasons, which is higher in magnitude as compared to observed cooling in summer and autumn seasons. This particular fact has led to a dominant warming trend by most of the station at annual time scale, which is higher in magnitude at high altitude stations mainly due to their dominated winter warming as compared to low altitude stations (Shrestha et al., 1999; Liu and Chen, 2000).

**Diurnal temperature range**

For the DTR, most of the stations show its drop throughout a year except during months of March and May, where particularly low altitude stations show its increase mainly due to higher warming in Tx than in Tn or higher cooling in Tn than in Tx (Table 4 and Fig. 4). Two stations (Chillas and Skardu) show a significant widening of DTR in May, followed by Chillas station in March, Deosai in August and Gupis in October months. Conversely, we observe high inter-station agreement of significant DTR decrease in September followed by in February. Such a trend is associated with the higher magnitude of cooling in Tx than in Tn (e.g. in September), cooling in Tx but warming in Tn or higher warming in Tn than in Tx (e.g. in February). We note that long term trends of increasing DTR throughout a year from low altitude stations (Table 6) are now mainly restricted to the period March-May, and within the months of October and December over the period 1995-2012. Within the rest of year, DTR has been decreasing since last two decades. Overall, high altitude stations exhibit though less strong but a robust pattern of year round significant decrease in DTR as compared to low altitude stations.

**Total precipitation**

We find that most of the stations show a clear signal of dryness during the period March-June, which is either relatively higher or similar at high altitude station than at low altitude stations (Table 5 and Fig. 4). During such period, significant drying is revealed by seven stations (Deosai, Dainyor, Yasin, Astore, Chillas, Gupis and Khunjrab) in March, by five stations (Dainyor, Rattu, Astore, Bunji and Chillas) in April, by two stations (Dainyor and Rattu) in May and by four stations (Dainyor, Rama, Rattu and Shigar) in June. We have observed similar significant drying during August by three stations (Rattu, Shigar and Gupis) and during October by three stations (Rattu, Shendure and Yasin). The Rattu station features a consistent drop in precipitation throughout a year except during the months of January and February where basically a neutral behavior is observed. Stations feature high agreement for an increase in precipitation during winter season (December to February) and during the month of September, where such increase is higher in magnitude at high altitude stations as compared to low altitude stations. We note that most of the stations within the UIB-West-upper region (monsoon dominated region) exhibit an increase in precipitation. Shendure, Yasin, Ziarat, Rattu, Shigar and Chillas are stations featuring significant increase in precipitation in either all or at least in one of the monsoon months. Such precise response of increase or decrease in precipitation at monthly scale is averaged out on a seasonal time scale, on which autumn and winter seasons show an increase while spring and summer seasons show a decrease. Annual trends in precipitation show a mixed response by roughly equal number of stations.

From our comparison of medium term trends at low altitude stations with their long term trends (See Table 5 and 6), we note that trends over the recent decades exhibit much higher magnitude of dryness during spring months, particularly for March and April, and of wetness particularly within the month of September – the last monsoonal month. Interestingly, shifts in the trend have been noticed during the summer months (June-August) where trends over recent decades exhibit drying but the long-term trends suggest wetter conditions. This may attribute to multi-decadal variability that is associated with the global indices, such as, NAO and ENSO, influencing the climatic processes over the region (Shaman and Tziperman, 2005; Syed et al., 2006). Only increase in September precipitation is consistent between the long-term trend and trend obtained over 1995-2012 at low altitude stations.

**Discharge**

Based on the median hydrograph of each stream flow gauge for the UIB (Fig. 3), we clearly show that both snow and glacier fed/melt regimes can be differentiated based on their runoff production time. Figure 3 suggests that Indus at Kharmong (Eastern UIB), Gilgit at Gilgit (Hindukush) and Astore at Doyian are primarily snow fed basins, generally featuring their peak runoff in July. The rest of the basins are mainly glacier fed basins that feature their peak runoff in August.

Based on 1995-2012 period, our trend analysis suggests an increase in discharge from most of the hydrometric stations within the UIB during October-June, which is higher in magnitude during May-June (Table 5). A discharge increase pattern seems to be more consistent with tendencies in the temperature record than in precipitation record. In contrast, most of the hydrometric stations experience a decreasing trend of discharge during the month of July, which is statistically significant out of five (Karakoram, Shigar, Shyok, UIB-Central and Indus at Kachura) regions, owing to drop in July temperatures. These regions, showing significant drop in discharge, are mainly high-altitude/latitude glacier-fed regions within the UIB. For August and September months, there is a mixed response, however, statistically significant trends suggest an increase in discharge out of two (Hindukush and UIB-West-lower) regions in August and out of four (Hindukush, western-Karakoram, UIB-West-lower and UIB-west) regions during September. We note that despite of the dominant cooling during September, discharge mainly drops during July, suggesting a strong impact of the cooling during such a month. Moreover, regions showing an increase in discharge during September are mainly the western region of UIB. Such an increase in discharge can mainly be attributed to increasing precipitation trends over such regions. Overall, discharge from UIB also decreases during the month of July, however, such a drop is not statistically significant. Possibly, the lack of statistical significance in the decrease of UIB discharge may possibly be due to integrating the response from its sub-regions, and a statistically significant signal might become apparent when looking at higher temporal resolution data, such as 10-day or 5-day average discharge. During winter, spring and autumn seasons, discharge at most sites increases while during summer season and on an annual time scale there is a mixed response.

Our long-term analysis reveals a rising trend of stream flow during the period (November to May) from most of the sites/regions (Table 6). Such rising trend is particularly higher in magnitude in May and also significant at relatively large number of gauging sites (14 among 16). In contrast to November-May period, there is a mixed signal of rising and falling stream flow trend among sites during June-October. The rising and falling stream flow trends at monthly time scale exhibit similar response when aggregated on a typical seasonal or annual time scales. Winter discharge features an increasing trend while for the rest of seasons and on an annual time scale, sites mostly exhibit a mixed response.

While comparing the long-term trends with the trends assessed from recent two decades, we note most prominent shifts in the sign of trends during the seasonal transitional month of June and within the high flow months July-September, which may attribute to higher summer cooling together with the enhanced precipitation under the influence of monsoonal precipitation regime in recent decades. For instance, long term trend suggests that discharge out of eastern-, central- and whole Karakoram, UIB-Central, Indus at Kachura, Indus at Partab Bridge and Astore regions is increasing while rest of regions feature a decreasing trend. However, trend from the recent two decades suggests the opposite sign of discharge coming out of such regions, except the regions of Astore, Hindukush, UIB-West-upper and its sub-regions, which consistently show similar sign of change. Such response may attribute to a multi-decadal variability of climatic processes over the region, which is driven by NAO and ENSO (Shaman and Tziperman, 2005; Syed et al., 2006).

## 5.2 Field significance of local trends and physical attribution

Based on number of local significant trends, we analyze their field significance for both positive and negative trends, separately (Table 7). We present the mean slope of the field significant local trends in order to present the dominant signal from the region. Our results show a unanimous field significant warming for most of regions in March followed by in August. Similarly, we generally find a field significant decrease in precipitation during month of March over all regions except Karakoram and UIB-Central regions. We find a field significant cooling over all regions during the months of July, September and October, which on a seasonal scale, dominates during autumn season followed by summer season. Interestingly, we note that most of the climatic trends are not field-significant during the transitional (or pre-monsoon) period of April-June. We found a general trend of narrowing DTR, which is associated with either warming of Tn against cooling of Tx or relatively lower cooling in Tn than in Tx. Field significant drying of the lower latitudinal regions (Astore, Himalaya, UIB-West-lower - generally snow-fed regions) is also observed particularly during the period March-September, thus for the spring and summer and for the annual time scale. On the other hand, we found an increasing (decreasing) trend in precipitation during winter and autumn (spring and summer) seasons for the Hindukush, UIB-West, UIB-West-upper and whole UIB while for the western Karakoram such increase in precipitation is observed during winter season only. For the whole Karakoram and UIB-central regions, field significant increase in precipitation is observed throughout a year except during the spring season where no signal is evident.

We have noted that for most of the regions the field significant cooling and warming trends are in good agreement against the trends in discharge from the corresponding regions. Such agreement is high for summer months, particularly for July and, during winter season, for the month of March. Few exceptions to such a consistency are the regions of Himalaya, UIB-West and UIB-West-lower, for which in spite of field significant cooling in month of July, discharge still features a positive trend. However, we note that the magnitude of increase in July discharge has substantially dropped when compared to the increase in previous (June) and following (August) months. Such a substantial drop in the July discharge increase rate is again consistent with the prevailing field significant cooling during July for the UIB-West and UIB-West-lower regions. Thus, the identified field significant climatic signals for the considered regions are further confirmed by their observed discharge tendencies. In case climatic trends are not field significant for a particular region, still trend in discharge out of that region represents its prevailing climatic state, since discharge is an integrated signal of controlling climatic variables.

Interestingly, we note that generally magnitude of cooling during September dominates the magnitude of cooling during July while magnitude of warming during March dominates the magnitude of warming during May. However, subsequent runoff response from the considered regions does not correspond with the magnitude of cooling and warming trends. In fact, most prominent increase in discharge is observed in May while decrease in July, suggesting them months of effective warming and cooling, respectively. Generally, periods of runoff decrease (in a sequence) span from May to September for the Karakoram, June to September for the UIB-Central, July to August for the western-Karakoram and UIB-West-upper, July to November for the Astore and only over July for the Hindukush and UIB regions. Regions of UIB-West-lower and Himalaya suggest decrease in discharge during months of April and February, respectively.

## 5.3 Tendencies versus latitude, longitude and altitude

In order to explore the geographical dependence of the climatic tendencies, we plot tendencies from the individual stations against their longitudinal, latitudinal and altitudinal coordinates (Figs. 5-7). We note that summer cooling is observed by all stations however stations between 75-76$^o$ E additionally show such cooling during the month of May in Tx, Tn and Tavg. Within 74-75$^o$ E, stations generally show a positive gradient towards west in terms of warming and cooling, particularly for Tn. DTR generally features a narrowing trend where magnitude of such a trend tends to be higher west of 75$^o$ longitude (Astore basin). Precipitation generally increases slightly but decreases substantially at 75$^o$ longitude. Discharge decreases at highest (UIB-east) and lowest (UIB-west) gauges in downstream order, while increases elsewhere.

Cooling or warming trends are much prominent at higher latitudinal stations, particularly for cooling in Tx and warming in Tn. Highest cooling and warming in Tavg is noted around 36$^o$N. Similarly, we have observed a highest cooling in Tx and warming in Tn, while Tx cooling dominates in magnitude as evident from Tavg. DTR generally tends to decrease towards higher latitudes where magnitude of decrease in a particular season/month is larger than increase in it for any other season/month. Highest increasing or decreasing trend in precipitation is observed below 36$^o$N where station below 35.5$^o$N show substantial decrease in annual precipitation mainly due to decrease in spring season and stations between 35.5-36°N show increase in annual precipitation mainly due to increase in winter precipitation.

Magnitude of cooling (warming) in Tn decreases (increases) at higher elevations. Stations below 3500 m asl feature relatively higher magnitude of cooling in Tx, which is also higher than warming trends in Tx as well as in Tn. Such signals are clear from tendencies in Tavg. Stations between the elevation range 2000-4000 m asl clearly show pronounced Tavg cooling than Tavg warming in certain months/seasons. For low-altitude stations and stations at highest elevation show the opposite response, featuring a pronounced warming in Tavg than its cooling in respective months/seasons. We note that precipitation trends from higher altitude stations are far more pronounced than in low altitude station, and clearly suggest drying of spring but wetting of winter seasons. Tendencies in DTR in high altitude stations are consistent qualitatively and quantitatively as compared to tendencies in low altitude stations.

## 6 Discussions

The hydrology of UIB dominates with the melt water runoff, which ensures the crucial water supply to the largest reservoir in Pakistan for reducing the ongoing electric shortfall by its use for hydro-power generation, and contributing to the economy through its use for mostly irrigated agricultural production downstream. The water availability from the UIB depends upon a highly seasonal moisture input from the distinct mode of large scale circulations; the summer monsoon system transporting moisture from the Bay of Bengal and Arabian Sea, and the westerly disturbances bringing moisture from the Mediterranean and Caspian Seas, to their far extremities over the region. An interaction among these large-scale circulations over the highly complex terrain of HKH within the UIB largely influences substantially its thermal regime, which in turn, is primarily responsible for the melt runoff generation. The extent of the existing permanent cryosphere within the UIB additionally influences the timings of melt runoff production and ensures to a certain extent the compensation for variability in the moisture input in a running or previous accumulation season. In view of the fact that reduction in snow amount is somewhat compensated by the glacier melt, one can expect little changes in the overall meltwater availability from the UIB during subsequent melt season. The reduction of snow, however, may affect the timing of water availability due to certain time delays associated with the migration of melting temperature up to the glaciated region.

In contrast, cooling tendencies during the melt season, even in the presence of abundant snow, may lead to both an overall decrease and delay in the melt runoff. Nevertheless, persistent changes in both can have strong impact on the long-term water balance of the study basin and subsequently the future water availability. Therefore, knowledge about the climatic regime prevailing over the UIB is utmost necessary for better management and use of available water resources in Pakistan at present and for the immediate revision of the near term future planning such as Water Vision 2025.

Earlier investigations of the UIB climatic regime have been mainly restricted to only a subset of six available low altitude, manual, valley-bottom stations, not fully representative of the active hydrologic regime of the UIB. For the first time, we present a comprehensive and systematic assessment of the climatic tendencies for two recent decades from the updated record of twelve high altitude automated weather stations from HKH ranges together with a full set of six low altitude stations, all covering the altitudinal range roughly between 1000 and 4500 m asl. First, we perform a quality control and homogeneity test, and then we correct the time series for its sequential dependence by removing the optimally identified lag-1 autocorrelation through an iterative procedure. We employed a widely used MK test for ensuring existence of a trend while true slope of a trend was estimated by the Sen's slope method on monthly to annual time scale. We have divided the UIB into pragmatic region of Astore, Gilgit, Hunza, Himalaya, Karakoram, UIB-Central, UIB-West, UIB-West-lower, UIB-West-upper and UIB itself depending upon available hydrometric station network, identified/known distinct hydrological regimes and in view of the existing topographic barriers of HKH massifs. Provided a particular region features more than one meteorological station, individual climatic trends within the region were tested for their field significance based upon number of positive/negative significant trends, which in turn compared with the trends of outlet discharge from the region in order to furnish physical attribution to statistically identified signal of change. We also compare results of our trend analysis, performed over the updated full length record from six low altitude stations (onward called as long term trend), with the reports from earlier studies analyzing only subset of these stations relatively over a shorter period.

**Cooling trends**

Our long term trend in Tavg suggests summer cooling at all stations which is mostly significant, while for autumn season and on an annual time scale we found a mixed response.

Comparing results of our updated analysis with Fowler and Archer (2005 and 2006), who have analyzed subset of low altitude stations for the period (1961-1999/2000), we found a qualitative agreement for summer cooling tendencies at Astore, Bunji, Gilgit and Skardu stations, and during autumn, only at Bunji station. Sheikh et al. (2009) have also reported cooling in the mean annual temperatures at Gilgit, Gupis and Bunji stations during the monsoon period (June-September). In contrast, autumn cooling at Gilgit station, winter cooling at two stations (Astore and Bunji) and spring and annual cooling at three stations (Astore, Bunji and Gilgit), reported in Fowler and Archer (2005 and 2006) are not consistent with our results, which suggest instead warming or no change. Such inconsistency is not assured at Bunji station as its winter cooling reported in Fowler and Archer (2005) is inconsistently reported as a warming trend in Fowler and Archer (2006), over the same period of record investigated. Sheikh et al. (2009) have reported cooling in mean annual temperatures over Gilgit, Gupis and Bunji stations. Our results of cooling in Tavg during the monsoon months are consistently observed for the neighboring regions, such as, Nepal, Himalayas (Sharma et al., 2000; Cook et al., 2003), northwest India (Kumar et al., 1994), Tibetan Plateau (Liu and Chen, 2000), central China (Hu et al., 2003), and central Asia (Briffa et al. 2001) for the respective investigated periods. For Tx, summer cooling tendencies at Astore, Bunji and Gilgit and autumn cooling at Bunji station are consistent with Fowler and Archer (2006). For Tn, our results are in high agreement for a significant summer and autumn cooling with Fowler and Archer (2006) and Khattak et al. (2011), and with the findings of an increasing snow cover extent for summer season as reported by Hasson et al. (2014b) over the region. Whereas, cooling tendencies during winter and spring seasons and on an annual time scale in all temperature variables (Fowler and Archer, 2005 and 2006; Khattak et al., 2011) instead have been inconsistently suggested either warming or no trend at all in our updated analysis. More surprisingly, Río et al. (2013) have reported overall warming trend over Pakistan (and UIB), at all timescales, which is in direct contrast with the cooling tendencies reported here and by the above mentioned studies regardless of the seasons.

We note that a robust pattern of long-term summer cooling in Tn, Tx and Tavg during June-October is weak over 1995-2012 period and has been restricted mainly to the monsoonal period of July-October, where cooling during months of July and September dominates in terms of magnitude. Cooling tendencies observed mostly during the monsoon season are attributed to coincident incursions of south Asian summer monsoon system and its precipitation (Cook et al., 2003) into the Karakoram through crossing Himalayas and within the UIB-West region for which the Himalayan barrier does not exist. Such phenomenon seems to be accelerated at present under the observed increasing trend in the cloud cover and in the number of wet days, particularly over the UIB-West (Bocchiola and Diolaiuti, 2013) and subsequently in the total amount of precipitation during the monsoon season. The enhanced monsoonal influence in the far north-west over the UIB-West region, and within the Karakoram, is consistent with the extension of the monsoonal domain northward and westward under the global warming scenario as projected by the multi-model mean from climate models participating in the Climate Model Intercomparison Project Phase 5 (CMIP5) (Hasson et al., 2015). Such hypothesis further needs a detailed investigation and it is beyond the scope of present study. Nevertheless, increasing cloud cover due to enhanced influence and frequent incursions of monsoonal system leads to reduction of incident downward radiations and results in cooling (or less warming) of Tx, which then under the clear sky conditions, continues as a result of evaporative cooling of the moisture-surplus surface under precipitation event (Wang et al., 2014) or due to irrigation (Kueppers et al., 2007). Han and Yang (2013) found irrigation expansion over Xinjiang, China as a major cause of observed cooling in Tavg, Tx and Tn during May-September over the period 1959-2006. Similar cloudy conditions most probably are mainly responsible for initially higher warming in Tn through blocking outgoing longwave radiations and creating a greenhouse effect, depending on the relative humidity conditions. Given that such cloudy conditions persist longer in time, Tx and Tn are more likely tend to cool. Yadav et al. (2004) have related the higher drop in minimum temperature to intense night time cooling of the deforested, thus moisture deficit, bare soil surface exposed to direct day time solar heating. Such an explanation is valid here only for the areas under deforestation and below the tree line.

**Warming trends**

Our findings of robust long term increasing trends in Tx and Tavg during November-May are consistent with the results from Khattak et al. (2011), who have analyzed data for the period 1967-2005. However, they have found highest rate of warming during winter season, instead we have found it during the spring season, which is consistent with findings of Sheikh et al. (2009) and Río et al. (2013). Our results of spring warming also agree well with the observation of a decreasing extent of spring snow cover worldwide and in the Northern Hemisphere over the period 1967 to 2012 (IPCC, 2013). Similarly, warming tendencies during winter at most of the stations are in good agreement with a decreasing snow cover extent over the study region during the period 2001-2012 (Hasson et al., 2014b). The long term warming tendencies (November to May) observed in the present study largely agree qualitatively with the findings of Fowler and Archer (2005 and 2006) for all temperature variables.

We have found the long term trend of winter warming in Tx at low altitude stations less significant during 1995-2012 accompanied by most of cooling tendencies during the months of February and December. Interestingly, well-agreed long-term cooling in Tx during June and warming during October are now featuring opposite signs of change by most of the low altitude stations. Similarly, long term warming trend in Tavg within November-May period has recently been restricted to mainly March-June period and within August and November months at low altitude stations, where most of these stations exhibit cooling tendencies during the winter months over the period 1995-2012. This suggests that a long-term trend of winter warming since 1961 (Fowler and Archer, 2006) is no more valid over 1995-2012 period.

Within the 1995-2012 period, our analysis suggests either cooling (or weaker warming) during the winter season both at low and high altitude stations, which is in direct contrast to the long term warming trends observed over the full length record (Fowler and Archer, 2005 and 2006; Sheikh et al., 2009; Khattak et al., 2011) at low altitude stations and particularly surprising given the observed winter warming worldwide. A recent shift of winter warming to cooling is however consistently observed over eastern United States, southern Canada and much of the northern Eurasia (Cohen et al., 2012). Such winter cooling is a result of falling tendency of winter time Arctic Oscillation, which partly driven dynamically by the anomalous increase in autumnal Eurasian snow cover (Cohen and Entekhabi, 1999), can solely explain largely the weakening (strengthening) of the westerlies (maridional flow) and favor anomalously cold winter temperatures and their falling trends (Thompson and Wallace, 1998 and 2001; Cohen et al., 2012). Weakening of westerlies during winter may explain an aspect of well agreed drying during subsequent spring season, and may further be associated with conditions more favorable for the monsoonal incursions from south into the UIB.

During the period 1995-2012, largely agreed warming in Tx dominates at low altitude stations as compared to high altitude stations, in contrast to warming in Tn, which is higher in magnitude among high altitude stations. Under the drying spring scenario, a less cloudy conditions associated with increasing number of dry days for the westerly precipitation regime (Hasson et al., 2015) are most probably responsible for warming in Tx, consistent with global warming signal. Trends in Tavg are dominated by trends in Tx during July-October while these are dominated by Tn, during rest of the year. Overall, trends based on recent two decades suggest higher magnitude of warming than the long term trends, which is consistent with the recent acceleration pattern of climatic changes (IPCC, 2013). Moreover, such warming tendencies (1995-2012), being restricted to months of March, May and November, relatively dominate in March at low altitude stations in terms of magnitude and significance but in May at high altitude stations in terms of magnitude only. Interestingly, a pronounced summer warming at higher elevations as reported in Tien Shan, central Asia (Aizen et al., 1997), over the Tibetan Plateau (Liu and Chen, 2000) and Nepal Himalayas (Shrestha et al., 1999), and as speculated for the UIB by Fowler and Archer (2006) by analyzing low altitude stations, is generally found invalid here. Instead of the summer warming, we have found higher rate of spring warming at higher altitude stations, which is again only valid for Tn.

Our results of long term increase in DTR at low altitude stations within the UIB are consistent with Fowler and Archer (2006), and over the India, with Kumar et al. (1994) and Yadav et al. (2004) but in direct contrast to decrease worldwide (Jones et al., 1999) and over northeast China (Wang et al., 2014). Contrary to the long term trends in DTR, trends over 1995-2012 period at low altitude stations show a decrease. Similarly, contrary to the reason of decrease in DTR worldwide and over northeast China (Jones et al., 1999; Wang et al., 2014), summer DTR decrease during 1995-2012 is attributed to stronger cooling in Tx than in Tn. The observed DTR increase during spring is attributed to stronger warming in Tx than in Tn, which is again contrary to the reason for DTR increase from the full length record over UIB and India (Fowler and Archer, 2006; Kumar et al., 1994; Yadav et al., 2004). It implies that though UIB features some common responses of trends in DTR when compared worldwide or to the neighbouring regions, however reasons of such common responses are still contradictory.

**Wetting and drying trends**

Khattak et al. (2011) have found no definite pattern of change in precipitation from the low altitude stations analyzed for the period 1967-2005. Similarly, Bocchiola and Diolaiuti (2013) report mostly not statistically significant changes in precipitation. From our long term precipitation analysis, we have found, a coherent (but again lacking statistical significance) pattern of change in precipitation, which indicates an increasing tendency during winter, summer and autumn seasons and on annual time scale, while a decreasing tendency during the spring months at most of the low altitude stations. Significant drying found at Bunji station during spring season is consistent with decreasing precipitation trend from Archer and Fowler (2004) during January-March period, while for Astore station such spring drying is consistent with their result of slight decrease in precipitation during April-June period. Our results of long term increasing trend in precipitation at Astore station for the winter, summer and autumn seasons is also consistent with Farhan et al. (2014).

We note that stations at high altitude suggest relatively enhanced monsoonal influence since six stations (Shendure, Yasin, Ziarat, Rattu and Chillas and Shigar) within the UIB-West and Central-Karakoram regions feature significant increase in precipitation in either all or at least one of the monsoon months. This is in good agreement with the projected intensification of south Asian summer monsoonal precipitation regime under enhanced greenhouse gas emission scenarios (Hasson et al., 2013, 2014a & 2015). At the low altitude stations, shifts of the long-term trends of increasing summer precipitation (June-August) to drying over the period 1995-2012 indicate a transition towards weaker monsoonal influence at lower levels. This may relate to the fact that the monsoonal currents crossing the western Himalayan barriers reach the central and western UIB at higher levels.

The precipitation increase during winter but decrease during spring season is associated with certain changes in the westerly precipitation regime under changing climate. For instance, spring drying is mainly consistent with the weakening and northward shift of the mid-latitude storm track (Bengtsson et al., 2006) and increase in the number of dry days within the westerly precipitation regime (Hasson et al., 2015). On the other hand, observed increase in the winter precipitation is consistent with the observations as well as future projections of more frequent incursions of the westerly disturbances into the region (Ridley et al., 2013; Cannon et al., 2015; Madhura et al., 2015), which together with drying of spring season, indicate less intermittent westerly precipitation regime in future, as reported by Hasson et al. (2015) based on CMIP5 climate models. In view of more frequent incursions of the monsoonal system and westerly disturbances expected in the future and certain changes projected for the overall seasonality/intermittency of their precipitation regimes by the climate models (Hasson et al., 2015), one expects changes in the time of the melt water availability from the UIB. Such hypothesis can be tested by assessing changes in the seasonality of precipitation and runoff based on observations analyzed here and also through modelling melt water runoff from the region under prevailing climatic conditions.

**Water availability**

Consistent with Khattak et al. (2011), our long term trend in summer season discharge suggests its increase for Indus at Kachura region while its decrease for UIB-West-upper and whole UIB regions, and also, an increase in the winter and spring discharges for all three regions. Observed increases in annual mean discharge from Astore basin for the full length of record and for the period 1995-2012 are consistent with findings from Farhan et al. (2014) for the period 1985-1995 and 1996-2010, respectively. Our long-term trend in Shigar discharge suggests partially consistent results with Mukhopadhyay et al. (2014) exhibiting its increase for June and August, however, in contrast, its slight decrease during July and September, though no trend was statistically significant. Moreover, Mukhopadhyay et al. (2014) have reported a downward trend of only June and July discharge after 2000. However, during the period 1995-2012, we have found a prominent drop in Shigar discharge for all four months June-September, which is higher in magnitude and statistically significant during July. We also found a change of sign in the long term discharge out of UIB-East over the period 1995-2012. Mukhopadhyay et al. (2014) related the drop in June and July months with drop in winter snow fall, which may only be partially true in view of relatively higher magnitude of drying in spring as observed in our analysis. Moreover, our analysis suggests that a recent drop in Shigar discharge is due to less snow amount available because of spring drying, an early snow melt under higher spring warming and concurrently less melting due to wide spread cooling during June-October, particularly at relevant (Shigar and Skardu) stations.

We note prominent shifts of long term trends of rising stream flow into falling during June-September over the period 1995-2012 for mostly the glacier-fed regions (Indus at Kachura, Indus at Partab Bridge, Eastern-, Central- and whole-Karakoram and UIB-Central), which may attribute to higher summer cooling together with certain changes in the precipitation regime during such period. Change in sign of discharge trend for the eastern-Karakoram (Shyok) is expected to substantially alter discharge at Kachura site, thus deriving a Shigar discharge by applying previously identified constant monthly fractions to the downstream Kachura gauge (Mukhopadhyay et al., 2014) would less likely yield a valid Shigar discharge for its period of missing record (1999-2010). Some regions, such as, UIB-West-upper and its sub-regions together with Astore basin and whole UIB are the regions consistently showing same sign of change in their long term trend when compared to the trends derived over the period 1995-2012.

During 1995-2012, the decreasing stream flow trend observed for mainly the glacier-fed regions is significant mostly during month of July. Despite the fact that cooling in July is less prominent than cooling in September over the period 1995-2012, it is much effective due to the fact that it coincides with the main glacial melt season. Such drop in July discharge, owing to decreased melting, results in reduced melt water availability but, at the same time, indicates positive basin storage, in view of enhanced moisture input. Similarly, increase in discharge during May and June is due to the observed warming, which though less prominent in magnitude than warming in March, is much effective since it coincides with the snow melt season. This suggests an early melt of snow and subsequently increased melt water availability, but concurrently, a lesser amount of snow available for the subsequent melt season. Such distinct changes in snow melt and glacier melt regimes are mainly due to non-uniform signs of change and magnitudes of trends in climatic variables at a sub-seasonal scale. This further emphasizes on a separate assessment of changes in both snow and glacier melt regimes, for which an adequate choice is the hydrological models that are able to distinctly simulate snow and glacier melt processes. Nevertheless, changes in both snow and glacier melt regimes all together can result in a sophisticated alteration of the hydrological regimes of UIB, requiring certain change in the operating curve of the Tarbela reservoir in future.

The discharge change pattern seems to be more consistent with tendencies in the temperature record than tendencies in the precipitation record. This points to the fact that the cryosphere melting processes are the dominating factor in determining the variability of the rivers discharge in the study region. However, changes in precipitation regime can still influence substantially the melt processes and subsequent meltwater availability. For instance, monsoon offshoots intruding into the region ironically result in declining river discharge (Archer, 2004), since such monsoonal incursions, crossing the Himalaya, mainly drop moisture over the high altitude regions and in the form of snow (Wake, 1989; Böhner, 2006). In that case, fresh snow and clouds firstly reduce the incident energy due to high albedo that results in immediate drop in the melt, and secondly, the fresh snow insulates the underlying glacier/ice, slowing down the whole melt process till earlier albedo rates are achieved. Thus, melting of the snow and glaciers and overall resultant meltwater availability is inversely correlated to number of snowfall events/days during the melt season (Wendler and Weller, 1974; Ohlendorf et al., 1997).

We note that certain combinations of months exhibit common responses, and that such combinations are different from those typically considered for averaging seasons such as MAM, JJA, SON and DJF. We, therefore, suggest that analysis must be performed using the highest available temporal resolution, because time averaging can mask important effects. We also emphasize that analysis merely based upon the typical seasons averages out the pivotal signal of change, which can only be clearly visible at fine temporal resolution. Trends for typical seasons are analyzed in the study merely for sake of comparing results with earlier studies.

In view of the sparse network of meteorological observations analyzed here, we need to clarify that the observed cooling and warming is only an aspect of the wide spread changes prevailing over the wide-extent UIB basin. This is much relevant for the UIB-Central region where we have only one station each from the eastern- and central- Karakoram (UIB-Central), which might not be representative exclusively for the hydro-climatic state over respective regions. Thus, field significant results for the whole Karakoram region are mainly dominated by contribution of relatively large number of stations within the western-Karakoram. Nevertheless, glaciological studies, reporting and supporting the Karakoram anomaly (Hewitt, 2005; Scherler et al., 2011; Bhambri et al., 2013) and possibly a non-negative mass balance of the aboded glaciers within eastern- and central-Karakoram (Gardelle et al., 2013), further reinforce our results. Moreover, our results agree remarkably well with the local narratives of climate change as reported by Gioli et al. (2013). Since the resultant aspect has been confirmed for the UIB and for its sub-regions to be significant statistically, and are further evident from the consistent runoff response and findings from the existing studies, we are confident that observed signal of hydroclimatic change dominates at the present at least qualitatively.

The hydro-climatic regime of UIB is substantially controlled by the interaction of large scale circulation modes and their associated precipitation regimes, which are in turn controlled by the global indices, such as, NAO and ENSO etc. Such phenomena need to be better investigated for in depth understanding of the present variability in the hydrological regime of UIB and for forecasting future changes in it. For future projections, global climate models at a broader scale and their downscaled experiments at regional to sub-regional scales are most vital datasets available, so far. However, a reliable future change assessment over the UIB from these climate models will largely depend upon their satisfactory representation of the prevailing climatic patterns and explanation of their teleconnections with the global indices, which are yet to be (fully) explored. The recent generations of the global climate models (CMIP5) feature various systematic biases (Hasson et al., 2013, 2014a and 2015) and exhibit diverse skill in adequately simulating prevailing climatic regimes over the region (Palazzi et al., 2014; Hasson et al., 2015). We deduce that realism of these climate models about the observed winter cooling over UIB much depends upon the reasonable explanation of autumnal Eurasian snow cover variability and its linkages with the large scale circulations (Cohen et al., 2012), while their ability to reproduce summer cooling signal is mainly restricted by substantial underestimation of the real extent of the south Asian summer monsoon owing to underrepresentation of High-Asian topographic features and absence of irrigation waters (Hasson et al., 2015). However, it is worth investigating data from high resolution Coordinated Downscaled Experiments (CORDEX) for South Asia for representation of the observed thermal and moisture regimes over the study region and whether such dynamically fine scale simulations feature an added value in their realism as compared to their forced CMIP5 models. Given these models do not adequately represent the summer and winter cooling and spring warming phenomena, we argue that modelling melt runoff under the future climate change scenarios as projected by these climate models is still not relevant for the UIB as stated by Hasson et al. (2014b). Moreover, it is not evident when the summer cooling phenomenon will end. Therefore, we encourage the impact assessment communities to model the melt runoff processes from the UIB, taking into account more broader spectrum of future climate change uncertainty, thus under both prevailing climatic regime as observed here and as projected by the climate models, considering them relevant for the short term and the long term future water availability, respectively.

## 7 Conclusions

The time period covered by our presented analysis is not long enough to disintegrate the natural variability such as ENSO signals from the transient climate change. Nevertheless, we assume that our findings supplement ongoing research on the question of dynamics of the existing water resources such as Karakoram Anomaly and the future water availability. In view of recently observed shifts and acceleration of the hydroclimatic trends over HKH ranges and within the UIB, we speculate an enhanced influence of the monsoonal system and its precipitation regime during the late-melt season. On the other hand, changes in the westerly disturbances and in the associated precipitation regime are expected to drive changes observed during winter, spring and early-melt season. The observed hydroclimatic trends, suggesting distinct changes within the period of mainly snow and glacier melt, indicate at present strengthening of the nival while suppression of the glacial melt regime, which all together will substantially alter the hydrology of UIB. However, such aspects need to be further investigated in detail by use of hydrological modelling, updated observations and relevant proxy datasets. The changes presented in the study earn vital importance when we consider the socio-economic effects of the environmental pressures. Reduction in melt water will result in limited water availability for the agricultural and power production downstream and may results in a shift in solo-season cropping pattern upstream. This emphasizes the necessary revision of WAPDA's near future plan i.e. Water Vision 2025 and recently released first climate change policy by the Government of Pakistan in order to address adequate water resources management and future planning in relevant direction. We summarize main findings of our study below:

- The common patterns of change ascertained are cooling during monsoon season and warming during pre-monsoonal or spring season. Pattern of tendencies derived for Tavg are more robust throughout a year as it is dominated by a relatively more robust pattern of cooling in Tx than in Tn, and similarly by a relatively more robust pattern of warming in Tn than in Tx. Such signal is averaged out in typical seasons and on annual time scale.

- The long-term summer cooling period of June-October has been shortened to July-October over the period 1995-2012 during which cooling becomes stronger, which further dominates during month of September followed by month of July in terms of higher magnitude and its statistical significance agreed among number of stations. Low and high altitude stations feature roughly similar magnitude of cooling during 1995-2012, which is however higher than the observed magnitude of warming in respective temperature variables during spring months.

- A strong long-term winter warming in Tx is either invalid or weaker over the period 1995-2012, which being restricted to March, May and November months, dominates during March and particularly higher at low altitude stations. Whereas long term warming in Tn is restricted during February-May and month of November, which dominates during March and February and prominent at higher altitude stations than low altitude stations.

- The long term trends of increasing DTR throughout a year at low altitude stations have been restricted mainly to March and May while for the rest of year, DTR has been decreasing over the period 1995-2012. Overall, high altitude stations exhibit though less strong but a robust pattern of significant decrease in DTR throughout a year as compared to low altitude stations.

- Long term summer precipitation increase shifts to drying over 1995-2012 period at low altitude stations, indicating a transition of the precipitation regime to weaker monsoonal influence at low altitudes. Over 1995-2012 period, well agreed increase (decrease) in precipitation for winter season and for month of September (March-June period) has been observed, which is higher in magnitude than the long term trends and also at high altitude stations as compared to low altitude stations. Six stations suggest a significant increase in monsoonal precipitation during all or at least one month.

- Long term discharge trends exhibit rising (falling) melt season runoff from regions of eastern-, central- and whole Karakoram, UIB-Central, Indus at Kachura, Indus at Partab Bridge and Astore (for rest of the regions). However, over the period 1995-2012 rising and falling discharge trends from respective regions show opposite behavior except for the Astore, Hindukush, UIB-West-upper and its sub-regions, which consistently show similar sign of change.

- Hydroclimatic trends are prominently distinct among certain time periods within a year rather than against their geographical distributions. However, high altitude data suggest more pronounced and updated signal of ongoing change.

- We have noted that for most of the regions the field significant cooling and warming trends are in good agreement against the trends in discharge from the region. Such agreement is high for summer months, particularly for July and, during winter season, for the month of March.

- Magnitude of subsequent runoff response from the considered regions does not correspond with the magnitude of climatic trends. In fact, most prominent increase is observed in May while decrease in July, suggesting them months of effective warming and cooling.

Table 1: Characteristics of the gauged and derived regions of UIB. Note: *Including nearby Skardu and Gilgit stations for the Karakoram and Deosai station for the UIB-Central regions

| S. No. | Watershed/ Tributary | Designated Discharge sites | Expression of Derived Discharge | Designated Name of the Region | Area (km$^2$) | Glacier Cover (km$^2$) | % Glacier Cover | % of UIB Glacier Aboded | Elevation Range (m) | Mean Discharge (m$^3$s$^{-1}$) | % of UIB Discharge | No of Met Stations |
|---|---|---|---|---|---|---|---|---|---|---|---|---|
| 1 | Indus | Kharmong | | UIB-East | 69,355 | 2,643 | 4 | 14 | 2250-7027 | 451 | 18.8 | 1 |
| 2 | Shyok | Yogo | | Eastern-Karakoram | 33,041 | 7,783 | 24 | 42 | 2389-7673 | 360 | 15.0 | 1 |
| 3 | Shigar | Shigar | | Central-Karakoram | 6,990 | 2,107 | 30 | 11 | 2189-8448 | 206 | 8.6 | 1 |
| 4 | Indus | Kachura | | Indus at Kachura | 113,035 | 12,397 | 11 | 68 | 2149-8448 | 1078 | 44.8 | |
| 5 | Hunza | Dainyor Bridge | | Western-Karakoram | 13,734 | 3,815 | 28 | 21 | 1420-7809 | 328 | 13.6 | 4 |
| 6 | Gilgit | Gilgit | | Hindukush | 12,078 | 818 | 7 | 4 | 1481-7134 | 289 | 12.0 | 5 |
| 7 | Gilgit | Alam Bridge | | UIB-West-upper | 27,035 | 4,676 | 21 | 25 | 1265-7809 | 631 | 27.0 | 9 |
| 8 | Indus | Partab Bridge | | Indus at Partab | 143,130 | 17,543 | 12 | 96 | 1246-8448 | 1788 | 74.3 | |
| 9 | Astore | Doyian | | Astore at Doyian | 3,903 | 527 | 14 | 3 | 1504-8069 | 139 | 5.8 | 3 |
| 10 | **UIB** | **Besham Qila** | | **UIB** | **163,528** | **18,340** | **11** | **100** | **569-8448** | **2405** | **100.0** | **18** |
| 11 | | | 4 – 2 – 1 | derived Shigar | | | | | | 305 | 12.7 | |
| 12 | | | 2 + 3 + 5 | Karakoram | 53,765 | 13,705 | 25 | 75 | 1420-8448 | 894 | 37.2 | *8 |
| 13 | | | 2 + 11 + 5 | derived Karakoram | | | | | | 993 | 41.3 | |
| 14 | | | 4 – 1 | UIB-Central | 43,680 | 9,890 | 23 | 54 | 2189-8448 | 627 | 26.1 | *4 |
| 15 | | | 10 – 4 | UIB-West | 50,500 | 5,817 | 13 | 32 | 569-7809 | 1327 | 55.2 | 14 |
| 16 | | | 10 – 4 – 7 | UIB-West-lower | 23,422 | 1,130 | 7 | 6 | 569-8069 | 696 | 28.9 | 5 |
| 17 | | | 1 + 16 | Himalaya | 92,777 | 3,773 | 5 | 20 | 569-8069 | 1147 | 47.7 | 7 |

Table 2: List of Meteorological Stations and their attributes. Inhomogeneity is found only in Tn over full period of record. Note: (*) represent inhomogeneity for 1995-2012 period only.

| S. | Station Name | Period From | Period To | Agency | Longitude | Latitude | Altitude | Inhomogeneity at |
|---|---|---|---|---|---|---|---|---|
| 1 | Chillas | 01/01/1962 | 12/31/2012 | PMD | 35.42 | 74.10 | 1251 | 2009/03 |
| 2 | Bunji | 01/01/1961 | 12/31/2012 | PMD | 35.67 | 74.63 | 1372 | 1977/11 |
| 3 | Skardu | 01/01/1961 | 12/31/2012 | PMD | 35.30 | 75.68 | 2210 | |
| 4 | Astore | 01/01/1962 | 12/31/2012 | PMD | 35.37 | 74.90 | 2168 | 1981/08 |
| 5 | Gilgit | 01/01/1960 | 12/31/2012 | PMD | 35.92 | 74.33 | 1460 | 2003/10* |
| 6 | Gupis | 01/01/1961 | 12/31/2010 | PMD | 36.17 | 73.40 | 2156 | 1988/12 1996/07* |
| 7 | Khunjrab | 01/01/1995 | 12/31/2012 | WAPDA | 36.84 | 75.42 | 4440 | |
| 8 | Naltar | 01/01/1995 | 12/31/2012 | WAPDA | 36.17 | 74.18 | 2898 | 2010/09* |
| 9 | Ramma | 01/01/1995 | 09/30/2012 | WAPDA | 35.36 | 74.81 | 3179 | |
| 10 | Rattu | 03/29/1995 | 03/16/2012 | WAPDA | 35.15 | 74.80 | 2718 | |
| 11 | Hushe | 01/01/1995 | 12/31/2012 | WAPDA | 35.42 | 76.37 | 3075 | |
| 12 | Ushkore | 01/01/1995 | 12/31/2012 | WAPDA | 36.05 | 73.39 | 3051 | |
| 13 | Yasin | 01/01/1995 | 10/06/2010 | WAPDA | 36.40 | 73.50 | 3280 | |
| 14 | Ziarat | 01/01/1995 | 12/31/2012 | WAPDA | 36.77 | 74.46 | 3020 | |
| 15 | Dainyor | 01/15/1997 | 07/31/2012 | WAPDA | 35.93 | 74.37 | 1479 | |
| 16 | Shendoor | 01/01/1995 | 12/28/2012 | WAPDA | 36.09 | 72.55 | 3712 | |
| 17 | Deosai | 08/17/1998 | 12/31/2011 | WAPDA | 35.09 | 75.54 | 4149 | |
| 18 | Shigar | 08/27/1996 | 12/31/2012 | WAPDA | 35.63 | 75.53 | 2367 | |

Table 3. List of SHP WAPDA Stream flow gauging stations in a downstream order along with their characteristics and period of record used. *Gauge is not operational after 2001.

| S. No. | Gauged River | Discharge Gauging Site | Period From | Period To | Degree Latitude | Degree Longitude | Height meters |
|---|---|---|---|---|---|---|---|
| 1 | Indus | Kharmong | May-82 | Dec-11 | 34.9333333 | 76.2166667 | 2542 |
| 2 | Shyok | Yogo | Jan-74 | Dec-11 | 35.1833333 | 76.1000000 | 2469 |
| 3 | Shigar | Shigar* | Jan-85 | Dec-01 | 35.3333333 | 75.7500000 | 2438 |
| 4 | Indus | Kachura | Jan-70 | Dec-11 | 35.4500000 | 75.4166667 | 2341 |
| 5 | Hunza | Dainyor | Jan-66 | Dec-11 | 35.9277778 | 74.3763889 | 1370 |
| 6 | Gilgit | Gilgit | Jan-70 | Dec-11 | 35.9263889 | 74.3069444 | 1430 |
| 7 | Gilgit | Alam Bridge | Jan-74 | Dec-12 | 35.7675000 | 74.5972222 | 1280 |
| 8 | Indus | Partab Bridge | Jan-62 | Dec-07 | 35.7305556 | 74.6222222 | 1250 |
| 9 | Astore | Doyian | Jan-74 | Aug-11 | 35.5450000 | 74.7041667 | 1583 |
| 10 | UIB | Besham Qila | Jan-69 | Dec-12 | 34.9241667 | 72.8819444 | 580 |

Table 4: Trend for Tx, Tn and DTR in °C yr⁻¹ (per unit time) at monthly to annual time scale over the period 1995-2012. Note: meteorological stations are ordered from top to bottom as highest to lowest altitude while hydrometric stations as upstream to downstream. Slopes significant at 90% level are given in bold while at 95% are given in bold and Italic. Color scale is distinct for each time scale where blue (red) refers to increasing (decreasing) trend

| Variable | Stations | Jan | Feb | Mar | Apr | May | Jun | Jul | Aug | Sep | Oct | Nov | Dec | DJF | MAM | JJA | SON | Ann. |
|---|---|---|---|---|---|---|---|---|---|---|---|---|---|---|---|---|---|---|
| Tx | Khunrab | 0.01 | -0.01 | 0.10 | 0.03 | 0.12 | -0.01 | -0.09 | 0.06 | -0.16 | 0.01 | 0.12 | 0.07 | 0.05 | 0.07 | -0.05 | 0.04 | 0.04 |
|  | Deosai | 0.02 | -0.05 | 0.07 | -0.01 | 0.06 | 0.01 | -0.19 | -0.01 | 0.00 | 0.02 | 0.06 | 0.05 | 0.08 | 0.06 | 0.03 | 0.02 | 0.06 |
|  | Shendure | -0.17 | **-0.09** | 0.01 | -0.03 | -0.06 | -0.10 | **-0.13** | ***-0.07*** | **-0.22** | -0.06 | 0.04 | -0.11 | -0.08 | -0.06 | **-0.11** | -0.05 | -0.05 |
|  | Yasin | 0.00 | -0.03 | **0.13** | -0.02 | 0.10 | 0.03 | -0.16 | -0.08 | **-0.35** | 0.12 | -0.02 | -0.10 | 0.03 | 0.08 | -0.06 | -0.01 | 0.05 |
|  | Rama | -0.06 | -0.07 | 0.02 | **-0.11** | 0.14 | 0.04 | -0.11 | -0.09 | **-0.29** | -0.10 | 0.01 | 0.00 | -0.04 | -0.04 | **-0.07** | -0.07 | ***-0.08*** |
|  | Hushe | -0.05 | -0.01 | 0.09 | 0.00 | 0.17 | -0.06 | -0.09 | 0.02 | **-0.20** | -0.09 | 0.01 | 0.03 | 0.02 | 0.03 | -0.02 | -0.03 | -0.03 |
|  | Ushkore | -0.04 | -0.02 | 0.10 | 0.03 | 0.25 | -0.01 | -0.12 | -0.06 | **-0.22** | -0.05 | 0.06 | -0.01 | 0.02 | 0.08 | -0.05 | -0.02 | -0.01 |
|  | Ziarat | 0.00 | -0.01 | 0.12 | -0.02 | 0.13 | 0.09 | -0.11 | -0.03 | **-0.21** | -0.04 | 0.09 | 0.04 | 0.06 | 0.06 | -0.02 | -0.04 | 0.01 |
|  | Naltar | -0.04 | -0.04 | 0.10 | -0.03 | 0.10 | 0.03 | -0.12 | -0.03 | **-0.19** | 0.03 | -0.01 | 0.01 | -0.02 | 0.07 | -0.03 | -0.05 | 0.00 |
|  | Rattu | ***-0.16*** | -0.10 | 0.04 | -0.03 | 0.11 | 0.14 | -0.06 | -0.05 | -0.17 | -0.23 | 0.04 | -0.15 | ***-0.12*** | -0.03 | 0.01 | -0.03 | **-0.07** |
|  | Shigar | -0.04 | -0.08 | -0.02 | -0.08 | **-0.38** | -0.15 | -0.08 | 0.03 | -0.01 | -0.09 | 0.11 | 0.01 | -0.02 | **-0.09** | **-0.09** | -0.02 | -0.02 |
|  | Skardu | 0.10 | 0.08 | 0.12 | 0.04 | 0.04 | -0.08 | **-0.10** | 0.06 | **-0.23** | -0.10 | -0.04 | -0.05 | -0.02 | 0.13 | -0.07 | **-0.09** | -0.02 |
|  | Astore | 0.09 | 0.00 | **0.20** | 0.03 | 0.18 | 0.06 | -0.05 | -0.03 | **-0.15** | -0.11 | 0.05 | 0.04 | 0.08 | 0.15 | -0.01 | -0.05 | 0.02 |
|  | Gupis | -0.05 | 0.03 | **0.27** | 0.11 | 0.20 | 0.01 | -0.09 | -0.13 | -0.09 | 0.12 | 0.12 | 0.03 | 0.11 | 0.20 | 0.03 | 0.03 | 0.07 |
|  | Dainyor | -0.04 | -0.08 | **0.23** | -0.02 | 0.15 | -0.19 | ***-0.18*** | 0.01 | -0.15 | -0.04 | 0.10 | -0.07 | **-0.06** | 0.14 | -0.08 | -0.01 | -0.02 |
|  | Gilgit | 0.09 | -0.07 | 0.12 | 0.03 | 0.15 | 0.02 | -0.15 | -0.08 | **-0.31** | -0.07 | 0.07 | -0.05 | -0.04 | 0.06 | -0.05 | **-0.08** | -0.05 |
|  | Bunji | 0.09 | -0.08 | 0.13 | 0.04 | 0.11 | 0.07 | -0.01 | 0.04 | **-0.22** | -0.12 | -0.01 | -0.08 | 0.00 | 0.11 | 0.02 | -0.07 | -0.02 |
|  | Chilas | 0.09 | -0.03 | **0.16** | 0.01 | 0.13 | 0.01 | **-0.15** | -0.06 | **-0.24** | 0.00 | 0.03 | -0.06 | -0.05 | 0.08 | **-0.07** | -0.05 | -0.06 |
| Tn | Khunrab | ***0.15*** | ***0.26*** | 0.16 | 0.03 | 0.18 | -0.02 | -0.04 | 0.00 | 0.01 | 0.05 | ***0.17*** | 0.10 | ***0.21*** | 0.08 | -0.01 | 0.06 | **0.09** |
|  | Deosai | 0.02 | 0.09 | **0.21** | 0.00 | 0.01 | 0.00 | 0.03 | -0.02 | -0.08 | 0.03 | 0.09 | 0.00 | 0.06 | 0.10 | -0.02 | 0.05 | **0.10** |
|  | Shendure | 0.04 | -0.03 | 0.10 | 0.06 | 0.05 | 0.00 | **-0.06** | 0.00 | -0.10 | -0.01 | 0.10 | 0.08 | 0.09 | 0.07 | -0.03 | 0.01 | 0.05 |
|  | Yasin | 0.09 | 0.07 | 0.12 | 0.02 | 0.10 | 0.01 | -0.11 | -0.05 | **-0.21** | 0.10 | 0.04 | -0.08 | 0.06 | **0.11** | -0.04 | 0.03 | ***0.08*** |
|  | Rama | -0.08 | 0.10 | 0.05 | 0.02 | 0.06 | 0.01 | 0.00 | 0.01 | -0.09 | 0.00 | 0.11 | 0.07 | -0.02 | 0.03 | 0.03 | 0.03 | 0.02 |
|  | Hushe | 0.00 | **0.14** | 0.08 | 0.02 | 0.14 | -0.04 | -0.08 | 0.04 | -0.09 | -0.04 | 0.04 | 0.01 | 0.06 | 0.06 | -0.01 | 0.01 | 0.01 |
|  | Ushkore | -0.06 | 0.05 | 0.08 | 0.09 | 0.13 | 0.00 | -0.04 | -0.02 | **-0.16** | -0.09 | 0.08 | 0.01 | 0.00 | 0.08 | 0.01 | -0.01 | 0.00 |
|  | Ziarat | 0.12 | ***0.23*** | 0.11 | 0.04 | 0.04 | 0.04 | -0.08 | 0.01 | -0.10 | -0.01 | 0.09 | 0.09 | ***0.17*** | 0.07 | 0.00 | 0.01 | ***0.06*** |
|  | Naltar | -0.01 | 0.08 | 0.10 | 0.02 | -0.01 | -0.03 | **-0.10** | -0.01 | **-0.07** | 0.00 | -0.03 | 0.00 | -0.07 | 0.10 | -0.03 | -0.01 | 0.04 |
|  | Rattu | -0.05 | 0.10 | -0.08 | -0.02 | 0.06 | 0.05 | -0.07 | 0.01 | -0.12 | -0.02 | 0.07 | 0.01 | 0.04 | -0.03 | 0.01 | **-0.08** | -0.04 |
|  | Shigar | 0.03 | 0.02 | -0.01 | -0.03 | **-0.21** | -0.09 | -0.07 | 0.05 | 0.07 | -0.11 | 0.05 | 0.04 | 0.01 | -0.02 | -0.06 | -0.01 | 0.01 |
|  | Skardu | -0.03 | 0.08 | -0.02 | -0.02 | -0.07 | -0.11 | -0.15 | -0.08 | -0.10 | ***-0.12*** | ***-0.14*** | -0.11 | ***-0.18*** | -0.01 | ***-0.12*** | ***-0.16*** | **-0.08** |
|  | Astore | 0.01 | 0.05 | 0.09 | 0.03 | -0.02 | 0.02 | -0.07 | 0.01 | **-0.10** | -0.05 | 0.05 | -0.08 | 0.06 | **0.11** | -0.01 | -0.03 | -0.02 |
|  | Gupis | -0.15 | -0.03 | **0.19** | 0.11 | 0.09 | 0.03 | -0.04 | 0.04 | -0.07 | -0.03 | -0.12 | -0.14 | -0.11 | 0.14 | -0.04 | -0.09 | 0.01 |
|  | Dainyor | -0.13 | 0.01 | ***0.13*** | 0.01 | 0.11 | -0.04 | ***-0.17*** | 0.03 | -0.06 | -0.02 | -0.06 | -0.05 | 0.01 | 0.07 | -0.03 | -0.04 | 0.01 |
|  | Gilgit | 0.03 | 0.10 | 0.06 | 0.04 | 0.04 | 0.05 | -0.01 | **0.26** | **0.30** | 0.05 | 0.09 | -0.01 | 0.08 | ***0.07*** | 0.06 | **0.19** | **0.08** |
|  | Bunji | 0.01 | 0.03 | 0.05 | 0.03 | 0.02 | 0.04 | -0.01 | 0.17 | 0.01 | 0.03 | **0.13** | 0.00 | 0.02 | 0.05 | 0.06 | 0.04 | 0.03 |
|  | Chilas | -0.09 | ***-0.18*** | 0.01 | -0.07 | 0.02 | -0.05 | -0.11 | -0.08 | **-0.21** | -0.10 | 0.00 | -0.06 | **-0.15** | -0.05 | **-0.07** | ***-0.11*** | -0.07 |
| DTR | Khunrab | -0.10 | ***-0.25*** | **-0.30** | **-0.19** | ***-0.24*** | -0.08 | ***-0.13*** | **-0.11** | **-0.11** | -0.04 | -0.03 | -0.05 | ***-0.17*** | -0.18 | -0.04 | ***-0.04*** | **-0.08** |
|  | Deosai | 0.07 | -0.09 | 0.01 | 0.11 | -0.05 | 0.05 | 0.16 | ***0.19*** | 0.01 | 0.02 | -0.01 | 0.03 | 0.01 | 0.00 | **0.13** | 0.01 | ***0.13*** |
|  | Shendure | -0.06 | -0.09 | -0.26 | ***-0.29*** | -0.17 | -0.08 | -0.03 | -0.05 | ***-0.09*** | -0.07 | -0.05 | -0.24 | -0.12 | ***-0.20*** | -0.10 | -0.06 | -0.15 |
|  | Yasin | -0.13 | ***-0.23*** | -0.05 | ***-0.15*** | -0.12 | ***-0.20*** | -0.13 | -0.11 | ***-0.22*** | ***-0.58*** | -0.24 | -0.19 | ***-0.08*** | -0.07 | ***-0.14*** | ***-0.25*** | ***-0.12*** |
|  | Rama | -0.05 | ***-0.16*** | -0.04 | ***-0.11*** | -0.04 | -0.02 | ***-0.15*** | ***-0.13*** | ***-0.27*** | -0.20 | -0.08 | -0.07 | ***-0.09*** | -0.07 | ***-0.07*** | ***-0.13*** | ***-0.08*** |
|  | Hushe | ***-0.08*** | ***-0.17*** | -0.01 | -0.05 | -0.02 | 0.00 | -0.03 | -0.02 | ***-0.07*** | 0.00 | -0.03 | -0.01 | ***-0.10*** | -0.01 | -0.02 | ***-0.03*** | ***-0.04*** |
|  | Ushkore | 0.00 | -0.06 | -0.02 | -0.08 | -0.01 | -0.05 | -0.01 | -0.02 | ***-0.08*** | -0.01 | -0.02 | -0.03 | -0.03 | -0.02 | -0.03 | ***-0.03*** | ***-0.03*** |
|  | Ziarat | -0.09 | ***-0.26*** | 0.02 | -0.02 | 0.01 | -0.01 | -0.05 | -0.01 | ***-0.10*** | -0.03 | -0.03 | -0.03 | -0.12 | -0.13 | 0.03 | -0.02 | ***-0.06*** |
|  | Naltar | -0.06 | ***-0.15*** | 0.02 | -0.06 | 0.06 | -0.02 | -0.02 | -0.02 | -0.09 | -0.03 | -0.03 | -0.13 | ***-0.08*** | 0.00 | -0.01 | -0.06 | ***-0.05*** |
|  | Rattu | -0.10 | ***-0.16*** | -0.04 | -0.10 | 0.02 | -0.04 | -0.09 | ***-0.11*** | ***-0.18*** | -0.16 | -0.18 | -0.15 | ***-0.12*** | -0.01 | -0.04 | -0.10 | ***-0.05*** |
|  | Shigar | 0.08 | 0.00 | -0.05 | 0.00 | 0.01 | 0.03 | -0.03 | -0.01 | -0.07 | 0.01 | **0.08** | 0.07 | 0.07 | 0.03 | -0.06 | 0.00 | -0.07 |
|  | Skardu | -0.04 | ***-0.14*** | 0.06 | 0.01 | **0.13** | 0.06 | -0.01 | -0.02 | ***-0.21*** | 0.04 | 0.03 | 0.14 | -0.07 | 0.07 | -0.01 | -0.01 | 0.00 |
|  | Astore | -0.02 | -0.13 | 0.13 | 0.00 | 0.05 | 0.00 | -0.03 | -0.07 | ***-0.08*** | 0.03 | -0.03 | 0.04 | -0.09 | 0.06 | -0.02 | -0.05 | -0.01 |
|  | Gupis | 0.04 | 0.00 | 0.15 | -0.01 | 0.10 | -0.01 | -0.03 | -0.10 | -0.05 | **0.16** | **0.16** | 0.15 | ***0.13*** | 0.07 | -0.06 | 0.09 | 0.09 |
|  | Dainyor | -0.05 | -0.09 | 0.06 | -0.11 | -0.21 | -0.19 | -0.11 | -0.07 | -0.10 | -0.44 | -0.01 | -0.07 | -0.09 | -0.07 | ***-0.23*** | -0.12 | ***-0.19*** |
|  | Gilgit | -0.13 | ***-0.19*** | 0.05 | -0.02 | 0.10 | -0.13 | ***-0.27*** | ***-0.26*** | ***-0.87*** | -0.18 | -0.09 | -0.02 | -0.11 | -0.03 | ***-0.15*** | ***-0.25*** | ***-0.18*** |
|  | Bunji | -0.04 | ***-0.14*** | 0.05 | 0.03 | 0.04 | -0.01 | -0.03 | -0.04 | ***-0.27*** | -0.03 | -0.16 | -0.10 | -0.07 | 0.06 | -0.01 | ***-0.14*** | -0.05 |
|  | Chilas | 0.07 | 0.09 | ***0.21*** | 0.11 | 0.13 | 0.03 | -0.04 | 0.04 | 0.00 | 0.08 | 0.01 | 0.04 | 0.10 | ***0.14*** | 0.02 | 0.02 | 0.02 |

Table 5: Same as Table 4 but trend slopes are for Tavg in °C yr⁻¹, for total P in mm yr⁻¹ and for mean Q in m³s⁻¹yr⁻¹. Color scale is distinct for each time scale where blue, yellow and orange (red, green and cyan) colors refer to decrease (increase) in Tavg, P and Q, respectively

| Variable | Stations | Jan | Feb | Mar | Apr | May | Jun | Jul | Aug | Sep | Oct | Nov | Dec | DJF | MAM | JJA | SON | Ann. |
|---|---|---|---|---|---|---|---|---|---|---|---|---|---|---|---|---|---|---|
| Tavg | Khunrab | 0.13 | 0.09 | 0.13 | 0.05 | 0.19 | 0.00 | -0.06 | 0.06 | -0.13 | 0.05 | *0.17* | 0.10 | *0.15* | 0.09 | -0.03 | 0.06 | 0.06 |
| | Deosai | 0.06 | 0.01 | **0.15** | 0.00 | 0.07 | 0.01 | -0.07 | 0.03 | -0.05 | 0.02 | 0.08 | 0.01 | *0.10* | 0.06 | 0.03 | 0.04 | **0.07** |
| | Shendure | -0.05 | -0.05 | 0.05 | 0.02 | 0.05 | -0.05 | -0.10 | -0.05 | -0.15 | -0.04 | 0.06 | -0.03 | 0.01 | -0.04 | -0.05 | -0.02 | 0.01 |
| | Yasin | 0.02 | 0.01 | *0.13* | 0.01 | 0.06 | 0.04 | -0.19 | -0.07 | -0.27 | 0.11 | 0.01 | 0.01 | 0.04 | *0.13* | -0.05 | 0.02 | 0.06 |
| | Rama | -0.12 | 0.02 | 0.05 | -0.06 | 0.07 | 0.01 | -0.03 | -0.03 | -0.19 | -0.09 | 0.05 | 0.02 | 0.02 | 0.00 | 0.00 | -0.01 | -0.04 |
| | Hushe | -0.03 | 0.05 | 0.06 | 0.02 | 0.14 | -0.05 | -0.07 | 0.02 | -0.13 | -0.07 | 0.03 | 0.04 | 0.01 | 0.06 | -0.01 | 0.00 | -0.01 |
| | Ushkore | -0.07 | 0.00 | 0.08 | 0.05 | *0.21* | 0.00 | -0.03 | -0.03 | -0.17 | -0.09 | 0.06 | 0.01 | 0.04 | 0.09 | -0.01 | -0.02 | 0.01 |
| | Ziarat | 0.04 | *0.11* | 0.10 | 0.02 | 0.09 | 0.06 | -0.09 | -0.03 | -0.15 | -0.03 | 0.09 | 0.03 | *0.08* | 0.07 | -0.02 | 0.00 | 0.05 |
| | Naltar | -0.03 | 0.01 | 0.08 | -0.05 | -0.11 | -0.07 | -0.12 | -0.06 | -0.17 | 0.00 | -0.03 | 0.01 | *-0.13* | 0.07 | -0.04 | -0.04 | 0.01 |
| | Rattu | -0.11 | -0.01 | -0.05 | -0.04 | 0.09 | 0.10 | -0.04 | 0.00 | -0.18 | -0.07 | 0.04 | -0.10 | -0.06 | 0.03 | 0.00 | -0.05 | -0.05 |
| | Shigar | 0.05 | -0.02 | 0.00 | -0.06 | *-0.30* | -0.13 | -0.13 | 0.04 | 0.04 | -0.14 | 0.07 | 0.03 | 0.01 | -0.04 | -0.07 | -0.01 | 0.00 |
| | Skardu | 0.02 | 0.11 | 0.07 | 0.01 | 0.02 | -0.10 | -0.15 | 0.04 | -0.17 | -0.11 | -0.06 | -0.07 | *-0.11* | 0.06 | -0.12 | -0.12 | -0.07 |
| | Astore | 0.10 | 0.03 | *0.12* | 0.01 | 0.13 | 0.03 | -0.05 | 0.00 | -0.14 | -0.09 | 0.03 | -0.01 | 0.05 | *0.13* | -0.02 | -0.03 | 0.01 |
| | Gupis | -0.08 | -0.06 | *0.22* | 0.09 | 0.13 | 0.00 | -0.05 | -0.05 | -0.08 | 0.06 | 0.04 | -0.07 | 0.02 | *0.14* | 0.02 | -0.01 | 0.03 |
| | Dainyor | -0.06 | -0.02 | *0.22* | -0.01 | *0.18* | -0.08 | -0.15 | 0.02 | -0.11 | -0.04 | -0.04 | -0.09 | -0.05 | *0.11* | -0.04 | -0.04 | 0.00 |
| | Gilgit | 0.02 | 0.01 | 0.11 | 0.03 | 0.06 | 0.04 | -0.06 | 0.05 | -0.09 | 0.00 | *0.08* | 0.05 | 0.03 | *0.08* | -0.02 | 0.00 | 0.03 |
| | Bunji | 0.06 | -0.02 | 0.06 | 0.02 | 0.05 | 0.02 | 0.00 | 0.09 | -0.07 | 0.03 | 0.06 | -0.06 | 0.03 | *0.08* | 0.06 | 0.00 | 0.01 |
| | Chilas | -0.02 | -0.14 | 0.06 | -0.02 | *0.16* | -0.03 | -0.12 | -0.07 | -0.19 | -0.07 | 0.01 | -0.06 | *-0.09* | 0.03 | -0.06 | -0.08 | -0.07 |
| P | Khunrab | *3.64* | 2.59 | -2.21 | -1.55 | -1.47 | 0.10 | 0.35 | 0.80 | 1.82 | -1.04 | 0.93 | 2.34 | *8.86* | -9.09 | -1.74 | 1.65 | 6.14 |
| | Deosai | 0.07 | 1.28 | **-1.42** | -0.66 | -1.27 | -0.89 | -0.40 | -1.00 | -0.77 | -0.42 | -0.81 | -0.32 | 1.40 | -4.50 | 0.00 | -1.99 | -7.87 |
| | Shendure | *1.54* | *2.75* | 1.35 | 2.13 | 0.60 | *2.12* | *1.83* | *1.38* | *1.45* | 1.24 | 1.40 | 2.12 | *5.71* | 4.50 | *4.82* | *3.58* | *29.53* |
| | Yasin | 1.33 | *1.86* | 0.59 | 0.25 | 1.22 | -0.50 | *1.45* | 0.02 | 0.92 | -0.21 | 0.06 | *2.74* | *6.09* | 0.60 | 1.32 | 0.26 | 11.70 |
| | Rama | 0.77 | 0.00 | **-6.50** | **-8.55** | **-4.52** | **-2.16** | **-2.35** | **-1.89** | **-1.44** | **-2.05** | **-3.74** | -2.03 | 7.00 | **-25.44** | **-8.41** | **-14.60** | **-43.92** |
| | Hushe | 0.65 | 0.24 | -1.23 | -0.30 | -1.97 | -1.21 | -1.71 | -0.60 | 0.73 | -0.64 | 0.11 | 0.72 | *3.47* | -4.51 | -4.28 | 0.70 | -5.54 |
| | Ushkore | 0.56 | -0.59 | **-2.33** | -1.02 | -1.97 | -0.93 | 0.00 | -0.09 | 1.01 | *-0.61* | -0.48 | 0.00 | -0.13 | -4.57 | -1.54 | -0.42 | -3.83 |
| | Ziarat | -0.91 | -0.56 | **-4.18** | **-5.28** | -1.83 | 0.25 | -0.67 | -0.18 | 1.20 | -0.58 | -0.43 | -0.61 | -3.59 | **-9.10** | -1.71 | -0.21 | **-16.32** |
| | Naltar | *3.75* | *8.41* | -4.49 | -0.36 | -2.75 | **-2.17** | 0.43 | **-2.33** | 1.32 | -0.36 | -0.70 | 1.35 | *19.43* | -8.39 | -0.99 | 2.42 | -0.28 |
| | Rattu | *1.36* | 2.13 | 0.08 | 0.36 | 0.26 | 0.53 | *0.91* | 0.75 | *0.95* | 0.84 | 0.69 | *1.53* | *4.43* | 1.23 | 1.81 | *2.36* | 10.64 |
| | Shigar | -0.24 | -0.89 | -1.07 | -2.62 | -2.05 | -0.33 | 1.75 | 0.80 | *2.40* | 1.13 | 0.18 | 1.49 | -1.67 | **-8.36** | 0.78 | *3.08* | -7.04 |
| | Skardu | -0.64 | 1.62 | 0.60 | 0.19 | -0.74 | -0.47 | -0.07 | -0.44 | 0.46 | 0.00 | 0.00 | 0.20 | 0.41 | 0.89 | -1.26 | 0.49 | 1.29 |
| | Astore | 0.00 | 0.41 | 0.12 | -1.41 | -0.48 | -0.16 | -0.08 | -0.29 | 0.57 | 0.00 | 0.00 | 0.29 | 1.50 | -1.36 | -1.63 | 0.34 | -0.16 |
| | Gupis | 0.65 | *0.97* | 0.81 | 0.38 | -0.06 | **-1.33** | -1.07 | -0.49 | 0.06 | 0.35 | 0.26 | *0.89* | 2.81 | 0.29 | **-3.49** | 0.43 | 4.46 |
| | Dainyor | -0.21 | 0.42 | *0.51* | 0.55 | 0.67 | *1.24* | 0.91 | -0.71 | -0.39 | 0.00 | 0.00 | 0.00 | *1.68* | 1.81 | 3.09 | -0.34 | 6.69 |
| | Gilgit | 0.98 | 0.45 | **-1.94** | -1.34 | -1.57 | -0.73 | 0.29 | **-3.99** | 0.32 | 0.00 | 0.00 | 0.30 | 0.00 | **-9.39** | **-9.60** | -0.92 | **-20.31** |
| | Bunji | 0.01 | -0.10 | **-1.06** | **-2.34** | 0.17 | 0.20 | -0.34 | -0.22 | 0.56 | -0.01 | 0.00 | 0.11 | -0.47 | -2.68 | -0.51 | 0.06 | 0.09 |
| | Chilas | 0.00 | 0.13 | -0.14 | -1.56 | 0.16 | 0.29 | -0.51 | 0.13 | *1.37* | -0.10 | 0.00 | 0.07 | 0.22 | -0.81 | -0.80 | 1.86 | 0.53 |
| Q | UIB-East | *-0.80* | 0.00 | 0.04 | 0.11 | -4.19 | 2.00 | -1.65 | 6.70 | -4.74 | -5.45 | *-2.46* | *-1.37* | *-0.75* | -2.64 | -2.62 | -0.86 | -1.73 |
| | Eastern-Karakoram | 0.06 | 0.08 | -0.10 | 0.00 | 1.96 | 0.96 | **-22.97** | 0.92 | -8.84 | -1.06 | 0.50 | -0.09 | 0.29 | 0.67 | 0.30 | -4.41 | -0.95 |
| | Central-Karakoram | 0.96 | 1.28 | 1.56 | -0.84 | 3.74 | -8.94 | **-37.93** | -9.08 | -5.98 | 0.71 | 2.50 | *2.76* | 1.13 | 1.13 | **-21.61** | 1.10 | -1.56 |
| | Kachura | 0.33 | 1.39 | *1.06* | -0.33 | -2.08 | -22.50 | **-50.04** | -16.74 | -4.25 | -2.18 | 0.59 | 2.64 | 0.46 | -0.81 | **-18.90** | -2.63 | -4.97 |
| | UIB-Central | *2.19* | 1.81 | 2.02 | -0.84 | 6.89 | -18.08 | **-43.79** | -20.20 | -4.88 | 1.05 | *4.38* | *2.34* | *2.00* | 1.79 | **-18.34** | 2.01 | -2.47 |
| | Western-Karakoram | 1.20 | 1.00 | 1.50 | *2.00* | 0.59 | 12.09 | -4.53 | -4.09 | *6.40* | *3.50* | *3.82* | *2.03* | *1.88* | 1.00 | -1.64 | *5.43* | 2.50 |
| | Karakoram | 1.88 | 2.00 | 1.33 | 1.00 | -5.82 | -7.80 | **-64.97** | -37.17 | -9.48 | 0.60 | *8.97* | *5.97* | 1.65 | 0.11 | **-24.43** | 5.64 | -3.90 |
| | Hindukush | 0.87 | 0.26 | 0.15 | 1.27 | 2.05 | 3.49 | -6.61 | *14.02* | *7.03* | 2.17 | 1.82 | 1.06 | 0.75 | 1.00 | 3.94 | 4.44 | 4.00 |
| | UIB-WU | 1.24 | *1.02* | 1.39 | *2.38* | *16.85* | 12.38 | **-25.48** | -15.50 | -1.28 | 0.69 | 0.98 | 0.52 | 0.55 | *7.76* | -3.68 | 0.45 | -1.25 |
| | Astore | 0.05 | 0.00 | 0.22 | 0.50 | 7.65 | 4.26 | -3.01 | 5.00 | -1.00 | -1.11 | -0.67 | 0.00 | 0.00 | 2.20 | 1.97 | -0.89 | 2.16 |
| | Partab_Bridge | 1.00 | -0.13 | 3.60 | 8.80 | *63.22* | -34.86 | **-39.86** | **-67.33** | 29.65 | 0.69 | *8.89* | *15.12* | *8.40* | *36.29* | **-67.00** | *9.81* | -12.40 |
| | UIB-WL | 1.88 | 0.41 | 6.39 | -0.52 | *41.58* | *59.50* | 28.19 | *81.58* | *30.99* | *16.18* | *5.17* | *2.33* | *1.92* | *19.90* | *65.53* | *16.02* | *25.44* |
| | UIB-WL-Partab | -3.00 | 0.80 | -4.38 | -0.82 | *87.89* | *51.53* | 9.00 | 17.67 | 2.71 | -12.24 | 1.40 | -6.00 | -3.74 | *28.32* | *47.93* | -3.00 | *18.94* |
| | UIB_West | 2.45 | 1.37 | 5.43 | 2.42 | *61.35* | *54.89* | 0.21 | *42.93* | *28.24* | *13.68* | *5.87* | 1.38 | 2.00 | *23.43* | *44.18* | *17.71* | *22.17* |
| | Himalaya | 0.30 | -0.32 | 4.10 | 0.91 | *43.99* | *62.23* | 12.43 | *83.33* | *22.43* | 9.97 | 2.32 | 0.23 | 1.17 | *26.64* | *57.88* | *7.75* | *24.66* |
| | UIB | 1.82 | *5.09* | 5.37 | -2.50 | 11.35 | 14.67 | **-46.60** | *41.71* | *35.22* | 10.17 | 5.29 | 0.75 | 1.91 | 15.72 | -1.40 | *19.35* | 4.25 |

Table 6: Results from low altitude stations for the full length of available record (as given in Table 2 and 3) for Tx, Tn, Tavg, DTR and P (rainfall) at monthly to annual time scales in respective units as per Table 4 and 5.

| Variable | Stations | Jan | Feb | Mar | Apr | May | Jun | Jul | Aug | Sep | Oct | Nov | Dec | DJF | MAM | JJA | SON | Ann. |
|---|---|---|---|---|---|---|---|---|---|---|---|---|---|---|---|---|---|---|
| Tx | Skardu | 0.07 | 0.06 | 0.06 | 0.05 | 0.07 | 0.02 | 0.01 | 0.00 | 0.02 | 0.03 | 0.06 | 0.06 | 0.05 | 0.07 | 0.01 | 0.04 | 0.04 |
|  | Astore | 0.02 | 0.01 | 0.06 | 0.04 | 0.05 | -0.01 | -0.01 | -0.02 | 0.00 | 0.02 | 0.03 | 0.04 | 0.02 | 0.06 | -0.01 | 0.02 | 0.02 |
|  | Gupis | 0.02 | 0.02 | 0.03 | 0.04 | 0.06 | -0.02 | -0.02 | -0.03 | -0.01 | 0.04 | 0.04 | 0.06 | 0.04 | 0.04 | -0.02 | 0.03 | 0.02 |
|  | Gilgit | 0.04 | 0.03 | 0.04 | 0.05 | 0.06 | -0.01 | -0.01 | -0.02 | -0.01 | 0.02 | 0.05 | 0.05 | 0.04 | 0.04 | -0.01 | 0.02 | 0.02 |
|  | Bunji | 0.02 | 0.01 | 0.04 | 0.00 | 0.01 | -0.06 | -0.05 | -0.05 | -0.04 | -0.04 | 0.03 | 0.02 | 0.02 | 0.02 | -0.05 | -0.02 | 0.00 |
|  | Chilas | -0.01 | -0.01 | 0.03 | 0.01 | 0.02 | -0.05 | -0.02 | -0.02 | -0.02 | 0.00 | 0.00 | 0.01 | 0.00 | 0.02 | -0.03 | 0.00 | 0.00 |
| Tn | Skardu | 0.00 | 0.02 | 0.00 | -0.01 | -0.01 | -0.04 | -0.04 | -0.04 | -0.04 | -0.05 | -0.02 | 0.01 | 0.01 | 0.00 | -0.04 | -0.04 | -0.02 |
|  | Astore | 0.02 | 0.01 | 0.03 | 0.03 | 0.04 | 0.00 | -0.02 | -0.02 | -0.01 | 0.00 | 0.02 | 0.01 | 0.01 | 0.04 | -0.01 | 0.01 | 0.01 |
|  | Gupis | -0.04 | -0.02 | -0.01 | -0.03 | -0.01 | -0.07 | -0.06 | -0.07 | -0.05 | -0.03 | -0.03 | -0.01 | -0.03 | -0.03 | -0.07 | -0.05 | -0.04 |
|  | Gilgit | 0.00 | 0.03 | 0.00 | -0.01 | 0.01 | -0.02 | -0.05 | -0.03 | -0.01 | -0.02 | -0.01 | 0.01 | 0.01 | 0.00 | -0.03 | -0.02 | -0.01 |
|  | Bunji | 0.01 | 0.01 | 0.03 | 0.00 | 0.00 | -0.03 | -0.04 | -0.03 | -0.03 | -0.03 | 0.00 | 0.01 | -0.01 | 0.01 | -0.04 | -0.04 | 0.00 |
|  | Chilas | 0.04 | 0.02 | 0.01 | 0.01 | 0.03 | -0.02 | -0.01 | -0.03 | -0.02 | 0.00 | 0.03 | 0.04 | 0.03 | 0.02 | -0.02 | 0.00 | 0.01 |
| Tavg | Skardu | 0.03 | 0.04 | 0.03 | 0.02 | 0.03 | -0.01 | -0.02 | -0.02 | -0.01 | 0.00 | 0.02 | 0.03 | 0.03 | 0.03 | -0.02 | 0.00 | 0.01 |
|  | Astore | 0.02 | 0.01 | 0.04 | 0.04 | 0.05 | 0.00 | -0.01 | -0.02 | 0.00 | 0.01 | 0.03 | 0.02 | 0.01 | 0.05 | -0.01 | 0.02 | 0.01 |
|  | Gupis | 0.00 | 0.00 | 0.00 | 0.00 | 0.03 | -0.04 | -0.05 | -0.05 | -0.03 | 0.00 | 0.01 | 0.02 | 0.00 | 0.01 | -0.04 | -0.01 | -0.01 |
|  | Gilgit | 0.02 | 0.03 | 0.02 | 0.02 | 0.04 | -0.02 | -0.03 | -0.03 | -0.02 | -0.01 | 0.03 | 0.03 | 0.03 | 0.02 | -0.03 | 0.00 | 0.00 |
|  | Bunji | 0.00 | 0.01 | 0.02 | -0.01 | -0.01 | -0.04 | -0.05 | -0.04 | -0.05 | -0.04 | 0.00 | 0.01 | 0.01 | 0.01 | -0.04 | -0.03 | 0.00 |
|  | Chilas | 0.02 | 0.00 | 0.01 | 0.01 | 0.03 | -0.03 | -0.02 | -0.02 | -0.02 | 0.00 | 0.02 | 0.02 | 0.01 | 0.02 | -0.03 | 0.00 | 0.00 |
| DTR | Skardu | 0.06 | 0.02 | 0.05 | 0.07 | 0.09 | 0.05 | 0.06 | 0.03 | 0.06 | 0.09 | 0.09 | 0.05 | 0.05 | 0.07 | 0.05 | 0.09 | 0.06 |
|  | Astore | 0.04 | 0.00 | 0.01 | 0.02 | 0.02 | -0.02 | 0.01 | 0.02 | 0.01 | 0.02 | 0.02 | 0.01 | 0.02 | 0.01 | 0.00 | 0.02 | 0.02 |
|  | Gupis | 0.08 | 0.06 | 0.05 | 0.07 | 0.09 | 0.06 | 0.06 | 0.04 | 0.07 | 0.10 | 0.09 | 0.08 | 0.09 | 0.06 | 0.05 | 0.08 | 0.07 |
|  | Gilgit | 0.04 | 0.02 | 0.04 | 0.07 | 0.06 | 0.00 | 0.05 | 0.04 | 0.05 | 0.05 | 0.07 | 0.05 | 0.04 | 0.04 | 0.03 | 0.06 | 0.04 |
|  | Bunji | 0.04 | 0.01 | 0.03 | 0.01 | 0.03 | 0.00 | -0.01 | 0.03 | 0.02 | 0.06 | 0.04 | 0.04 | 0.02 | 0.04 | 0.00 | 0.03 | 0.02 |
|  | Chilas | -0.04 | -0.02 | 0.00 | 0.00 | 0.00 | -0.03 | -0.01 | 0.01 | 0.01 | -0.01 | -0.02 | -0.03 | -0.03 | 0.00 | -0.01 | -0.01 | -0.02 |
| P | Skardu | 0.30 | 0.32 | 0.16 | 0.16 | -0.02 | 0.08 | 0.06 | 0.19 | 0.07 | 0.00 | 0.00 | 0.15 | 0.98 | 0.45 | 0.29 | 0.12 | 1.76 |
|  | Astore | 0.00 | -0.28 | -0.78 | -0.51 | -0.25 | 0.27 | 0.19 | 0.06 | 0.02 | -0.05 | 0.02 | -0.08 | 0.24 | -1.31 | 0.45 | 0.06 | -1.33 |
|  | Gupis | 0.08 | 0.04 | 0.28 | 0.30 | -0.08 | 0.00 | 0.24 | 0.18 | 0.00 | 0.00 | 0.00 | 0.00 | 0.11 | 0.20 | 0.32 | -0.09 | 2.00 |
|  | Gilgit | 0.00 | 0.00 | -0.02 | 0.05 | -0.05 | 0.23 | 0.01 | 0.01 | 0.03 | 0.00 | 0.00 | 0.00 | 0.02 | -0.44 | 0.28 | 0.10 | 0.38 |
|  | Bunji | 0.00 | -0.06 | -0.14 | 0.02 | -0.17 | 0.09 | 0.05 | 0.12 | 0.11 | -0.03 | 0.00 | 0.00 | 0.13 | -0.59 | 0.36 | 0.09 | 0.21 |
|  | Chilas | 0.00 | 0.04 | -0.12 | 0.00 | -0.01 | 0.10 | 0.07 | 0.07 | 0.07 | -0.02 | 0.00 | 0.00 | 0.25 | -0.12 | 0.51 | 0.03 | 0.70 |
| Q | UIB-East | 0.58 | 0.89 | 1.18 | 0.80 | 0.08 | -12.94 | -21.37 | -10.53 | -1.42 | -0.18 | 0.06 | 0.16 | 0.55 | 1.10 | -14.86 | -0.57 | -1.59 |
|  | Eastern-Karakoram | 0.00 | 0.00 | -0.04 | -0.08 | 1.79 | 6.46 | 5.17 | 6.81 | 4.34 | 1.31 | 0.24 | 0.00 | 0.07 | 0.41 | 7.08 | 2.05 | 2.43 |
|  | Central-Karakoram | 0.32 | -0.07 | -0.51 | -0.67 | 6.13 | 3.85 | -1.22 | 6.30 | -7.40 | -4.08 | -1.36 | -0.29 | -0.35 | 1.75 | 6.22 | -2.80 | 0.31 |
|  | Kachura | 1.04 | 1.40 | 1.19 | 0.43 | 6.06 | 12.88 | 14.75 | 19.45 | 14.27 | 3.69 | 1.14 | 1.13 | 1.12 | 2.67 | 19.20 | 6.12 | 7.19 |
|  | UIB-Central | 0.35 | 0.21 | -0.19 | -0.43 | 9.99 | 20.49 | 13.74 | 20.73 | -4.95 | -2.15 | -0.80 | -0.29 | -0.30 | 2.76 | 17.69 | -2.84 | 3.30 |
|  | Western-Karakoram | 0.04 | 0.00 | 0.00 | 0.00 | 0.29 | -3.75 | -12.69 | -13.75 | -2.14 | -0.24 | 0.18 | 0.20 | 0.13 | 0.24 | -10.23 | -0.59 | -2.55 |
|  | Karakoram | 0.28 | -0.20 | -0.60 | 0.33 | 9.67 | 24.33 | 8.29 | 8.13 | -7.57 | -2.18 | -0.59 | 0.63 | -0.15 | 4.17 | 24.39 | -4.36 | 6.44 |
|  | Hindukush | 0.00 | 0.05 | 0.04 | 0.19 | 3.31 | -1.00 | -0.85 | 0.11 | 0.64 | 0.23 | 0.15 | 0.13 | 0.04 | 1.25 | 0.24 | 0.31 | 0.48 |
|  | UIB-WU | 0.58 | 0.60 | 0.33 | 0.51 | 3.55 | -1.86 | -12.74 | -12.50 | 0.68 | 1.48 | 1.02 | 0.71 | 0.48 | 1.30 | -6.83 | 1.22 | -0.95 |
|  | Astore | 0.28 | 0.24 | 0.32 | 0.97 | 3.52 | 1.29 | -0.62 | 0.54 | 0.16 | 0.28 | 0.32 | 0.23 | 0.31 | 1.63 | 0.43 | 0.28 | 0.76 |
|  | Partab_Bridge | 1.01 | 0.49 | 0.44 | 1.93 | 18.03 | 13.07 | 12.89 | -8.37 | 9.74 | 3.84 | 2.61 | 1.63 | 1.74 | 6.84 | 7.05 | 4.93 | 4.72 |
|  | UIB-WL | 1.94 | 1.96 | 3.49 | 0.17 | 2.89 | -12.90 | -25.95 | -12.06 | -1.35 | 1.57 | 1.94 | 2.35 | 1.92 | 1.93 | -13.82 | 0.48 | -2.63 |
|  | UIB-WL-Partab | 1.58 | 1.87 | 2.11 | -0.82 | -0.30 | -22.26 | -16.35 | -17.07 | 0.02 | -2.20 | 0.23 | 1.18 | 1.32 | 0.34 | -22.10 | -0.99 | -5.40 |
|  | UIB_West | 2.02 | 2.01 | 2.73 | 1.12 | 8.00 | -19.88 | -32.88 | -23.24 | -5.13 | 1.95 | 2.59 | 2.40 | 2.18 | 3.99 | -25.21 | 0.93 | -4.03 |
|  | Himalaya | 3.23 | 3.91 | 4.73 | 2.33 | -0.33 | -32.29 | -69.33 | -17.55 | -4.61 | -0.05 | 3.40 | 2.05 | 3.37 | 6.86 | -40.09 | -0.72 | -6.13 |
|  | UIB | 3.00 | 3.33 | 3.53 | 0.62 | 12.97 | -8.84 | -13.31 | -3.24 | 8.19 | 4.03 | 3.92 | 3.04 | 3.04 | 5.00 | -6.15 | 5.14 | 2.23 |

Table 7: Field significance of the climatic trends for all regions considered along with trend in their Q at monthly to annual time scales over the period 1995-2012. Color scale as in Table 5

| Regions | Variables | Jan | Feb | Mar | Apr | May | Jun | Jul | Aug | Sep | Oct | Nov | Dec | DJF | MAM | JJA | SON | Ann. |
|---|---|---|---|---|---|---|---|---|---|---|---|---|---|---|---|---|---|---|
| **Astore** | Tx | -0.17 | | | | | | | | -0.21 | | | -0.42 | -0.16 | | | | -0.06 |
| | Tn | | | | | | | -0.10 | | | -0.10 | -0.12 | | | | | -0.10 | |
| | Tavg | -0.15 | | | | | | -0.13 | | | -0.21 | | | | | | | -0.05 |
| | DTR | | -0.22 | | | | | | | -0.13 | | | -0.17 | -0.07 | | | -0.06 | -0.08 |
| | P | | | -3.73 | -7.50 | -4.60 | -2.18 | -1.90 | -1.80 | -2.11 | | | | | -19.25 | -6.02 | -18.93 | -38.01 |
| | Q | 0.05 | 0.00 | 0.22 | 0.50 | 7.65 | 4.26 | -3.01 | 5.00 | -1.00 | -1.11 | -0.67 | 0.00 | 0.00 | 2.20 | 1.97 | -0.89 | 2.16 |
| **Hindukush** | Tx | | -0.11 | 0.23 | | | | -0.19 | | -0.29 | | | -0.18 | | | | -0.12 | -0.09 |
| | Tn | | | | | | | | 0.25 | 0.24 | | -0.18 | -0.24 | | | 0.09 | 0.10 | |
| | Tavg | | | 0.18 | | | | -0.11 | 0.08 | -0.25 | | | -0.13 | | | | -0.10 | |
| | DTR | -0.21 | | -0.11 | -0.18 | -0.25 | -0.28 | -0.19 | -0.36 | -0.40 | -0.52 | -0.38 | | 0.03 | -0.16 | -0.18 | -0.33 | -0.20 |
| | P | 1.30 | | -1.94 | | | | | 1.00 | | 1.05 | 0.31 | | 1.31 | 4.73 | -10.19 | 2.39 | |
| | Q | 0.87 | 0.26 | 0.15 | 1.27 | 2.05 | 3.49 | -6.61 | 14.02 | 7.03 | 2.17 | 1.82 | 1.06 | 0.75 | 1.00 | 3.94 | 4.44 | 4.00 |
| **Himalaya** | Tx | -0.17 | -0.10 | | | | | -0.22 | | -0.21 | -0.19 | | -0.28 | -0.16 | | -0.07 | -0.12 | -0.06 |
| | Tn | | -0.23 | 0.26 | | | -0.14 | -0.15 | 0.18 | | -0.16 | -0.18 | -0.14 | -0.18 | | -0.13 | -0.14 | 0.02 |
| | Tavg | -0.15 | | 0.25 | | | | -0.18 | 0.17 | -0.18 | -0.18 | -0.09 | -0.08 | -0.11 | | -0.10 | -0.13 | -0.07 |
| | DTR | -0.02 | -0.20 | 0.18 | -0.18 | | | -0.13 | -0.18 | -0.36 | -0.25 | | | -0.12 | | -0.08 | -0.19 | -0.09 |
| | P | | | -2.29 | -5.71 | -4.60 | -2.18 | -1.90 | -1.80 | -2.11 | | | 0.42 | | -12.15 | -6.02 | -18.93 | -38.01 |
| | Q | 0.30 | -0.32 | 4.10 | 0.91 | 43.99 | 62.23 | 12.43 | 83.33 | 22.43 | 9.97 | 2.32 | 0.23 | 1.17 | 26.64 | 57.88 | 7.75 | 24.66 |
| **West Karakoram** | Tx | | | 0.23 | | | | -0.18 | | -0.17 | -0.16 | | -0.06 | | | | | |
| | Tn | | 0.22 | 0.13 | | | | -0.13 | | | | | | 0.17 | | | | 0.05 |
| | Tavg | -0.15 | | 0.22 | -0.09 | | | -0.14 | | -0.15 | | | | | | | | |
| | DTR | | -0.22 | | | | | | | -0.13 | | | -0.17 | -0.07 | | | -0.06 | -0.08 |
| | P | | | | | 1.17 | 1.09 | | | | | | 3.81 | 9.08 | | | | |
| | Q | 1.20 | 1.00 | 1.50 | 2.00 | 0.59 | 12.09 | -4.53 | -4.09 | 6.40 | 3.50 | 3.82 | 2.03 | 1.88 | 1.00 | -1.64 | 5.43 | 2.50 |
| **Karakoram** | Tx | | -0.11 | 0.23 | | | | -0.18 | | -0.22 | -0.16 | | -0.06 | | | | -0.12 | -0.06 |
| | Tn | | -0.11 | 0.23 | | | | -0.18 | | -0.22 | -0.16 | | -0.06 | | | | -0.12 | -0.06 |
| | Tavg | | 0.22 | 0.13 | | | -0.14 | -0.14 | 0.25 | 0.46 | -0.16 | -0.14 | -0.18 | -0.16 | 0.17 | -0.08 | 0.06 | -0.05 |
| | DTR | -0.15 | | 0.22 | -0.09 | | | -0.15 | 0.08 | -0.16 | -0.12 | -0.09 | | | | -0.13 | -0.14 | -0.08 |
| | P | | 2.95 | 1.97 | | 1.17 | 1.72 | | 1.58 | 2.15 | 1.43 | 2.40 | 2.69 | 6.39 | | 5.39 | 5.76 | 45.07 |
| | Q | 1.88 | 2.00 | 1.33 | 1.00 | -5.82 | -7.80 | -64.97 | -37.17 | -9.48 | 0.60 | 8.97 | 5.97 | 1.65 | 0.11 | -24.43 | 5.64 | -3.90 |
| **UIB Central** | Tx | | | | | | | -0.26 | | -0.20 | -0.16 | | | | | | -0.12 | |
| | Tn | | | 0.26 | | | -0.14 | -0.20 | | | -0.16 | -0.18 | -0.16 | | | -0.17 | -0.18 | 0.02 |
| | Tavg | | | 0.25 | | | | -0.20 | | | -0.18 | -0.15 | -0.09 | | | -0.13 | -0.14 | -0.08 |
| | DTR | 0.13 | | | | | | | | | | | 0.09 | | | | | |
| | P | | 2.95 | 1.97 | | | 2.35 | | 1.58 | 2.15 | 1.43 | 2.40 | 1.57 | 5.99 | | 5.39 | 5.76 | 45.07 |
| | Q | 2.19 | 1.81 | 2.02 | -0.84 | 6.89 | -18.08 | -43.79 | -20.20 | -4.88 | 1.05 | 4.38 | 2.34 | 2.00 | 1.79 | -18.34 | 2.01 | -2.47 |
| **UIB** | Tx | -0.14 | -0.11 | 0.40 | | | | -0.20 | | -0.22 | -0.20 | | -0.25 | | | -0.09 | -0.12 | -0.09 |
| | Tn | | 0.49 | 0.38 | | | | -0.13 | 0.31 | | | -0.17 | | | | 0.37 | -0.14 | 0.27 |
| | Tavg | | | 0.37 | | | | -0.15 | 0.13 | -0.19 | -0.16 | | -0.11 | | | -0.10 | -0.12 | -0.08 |
| | DTR | | -0.19 | | -0.14 | | | -0.17 | -0.24 | -0.25 | -0.38 | | 0.11 | -0.13 | -0.10 | -0.17 | -0.09 | |
| | P | | | -2.17 | | 1.17 | -1.42 | | -2.40 | 1.65 | 1.10 | | 1.97 | 5.98 | -11.49 | -7.91 | 3.68 | |
| | Q | 1.82 | 5.09 | 5.37 | -2.50 | 11.35 | 14.67 | -46.60 | 41.71 | 35.22 | 10.17 | 5.29 | 0.75 | 1.91 | 15.72 | -1.40 | 19.35 | 4.25 |
| **UIB West** | Tx | -0.14 | -0.11 | 0.23 | | | | -0.18 | | -0.22 | -0.21 | | -0.25 | -0.11 | | -0.09 | -0.12 | -0.10 |
| | Tn | | | | | | | -0.12 | 0.22 | | | -0.18 | | | | | -0.13 | |
| | Tavg | -0.15 | | 0.20 | | | | -0.13 | 0.13 | -0.19 | -0.19 | | -0.11 | | | -0.11 | | -0.07 |
| | DTR | -0.18 | -0.20 | -0.10 | -0.16 | | | -0.17 | -0.24 | -0.27 | -0.38 | | | 1.71 | -0.10 | -0.13 | -0.10 | -0.19 | -0.10 |
| | P | | | -2.17 | -5.71 | 1.17 | | | -2.40 | | | | | 6.90 | -11.49 | -7.91 | 2.63 | |
| | Q | 2.45 | 1.37 | 5.43 | 2.42 | 61.35 | 54.89 | 0.21 | 42.93 | 28.24 | 13.68 | 5.87 | 1.38 | 2.00 | 23.43 | 44.18 | 17.71 | 22.17 |
| **UIB West Lower** | Tx | -0.17 | -0.10 | | | | | -0.16 | | -0.21 | -0.20 | | -0.28 | -0.16 | | -0.07 | -0.13 | -0.06 |
| | Tn | | -0.23 | | | | | -0.10 | 0.18 | | | -0.12 | -0.18 | | | -0.08 | -0.12 | |
| | Tavg | -0.15 | | | | | | -0.13 | 0.17 | | -0.19 | | -0.07 | -0.11 | | -0.06 | -0.11 | -0.07 |
| | DTR | -0.15 | -0.20 | 0.18 | -0.18 | | | -0.13 | -0.18 | -0.36 | -0.25 | | | -0.12 | | -0.08 | -0.19 | -0.09 |
| | P | | | -2.29 | -5.71 | -4.60 | -2.18 | -1.90 | -1.80 | -2.11 | | | 0.42 | | -12.15 | -6.02 | -18.93 | -38.01 |
| | Q | 1.88 | 0.41 | 6.39 | -0.52 | 41.58 | 59.50 | 28.19 | 81.58 | 30.99 | 16.18 | 5.17 | 2.33 | 1.92 | 19.90 | 65.53 | 16.02 | 25.44 |
| **UIB West Upper** | Tx | -0.14 | -0.11 | 0.23 | | | | -0.18 | | -0.22 | -0.21 | | -0.25 | -0.11 | | -0.09 | -0.12 | -0.10 |
| | Tn | | 0.22 | 0.13 | | | | -0.13 | 0.25 | 0.24 | | -0.18 | -0.24 | 0.17 | | 0.09 | 0.10 | 0.05 |
| | Tavg | -0.15 | | 0.20 | -0.09 | | | -0.13 | 0.08 | -0.20 | | | -0.13 | | | | -0.10 | |
| | DTR | -0.21 | -0.22 | -0.11 | -0.18 | -0.25 | -0.28 | -0.19 | -0.36 | -0.28 | -0.52 | -0.38 | -0.17 | 0.06 | -0.16 | -0.11 | -0.19 | -0.11 |
| | P | 1.30 | | -1.94 | | 1.17 | 1.09 | 1.00 | | 1.40 | 0.31 | | 2.14 | 6.90 | -10.19 | -9.80 | 2.63 | |
| | Q | 1.24 | 1.02 | 1.39 | 2.38 | 16.85 | 12.38 | -25.48 | -15.50 | -1.28 | 0.69 | 0.98 | 0.52 | 0.55 | 7.76 | -3.68 | 0.45 | -1.25 |

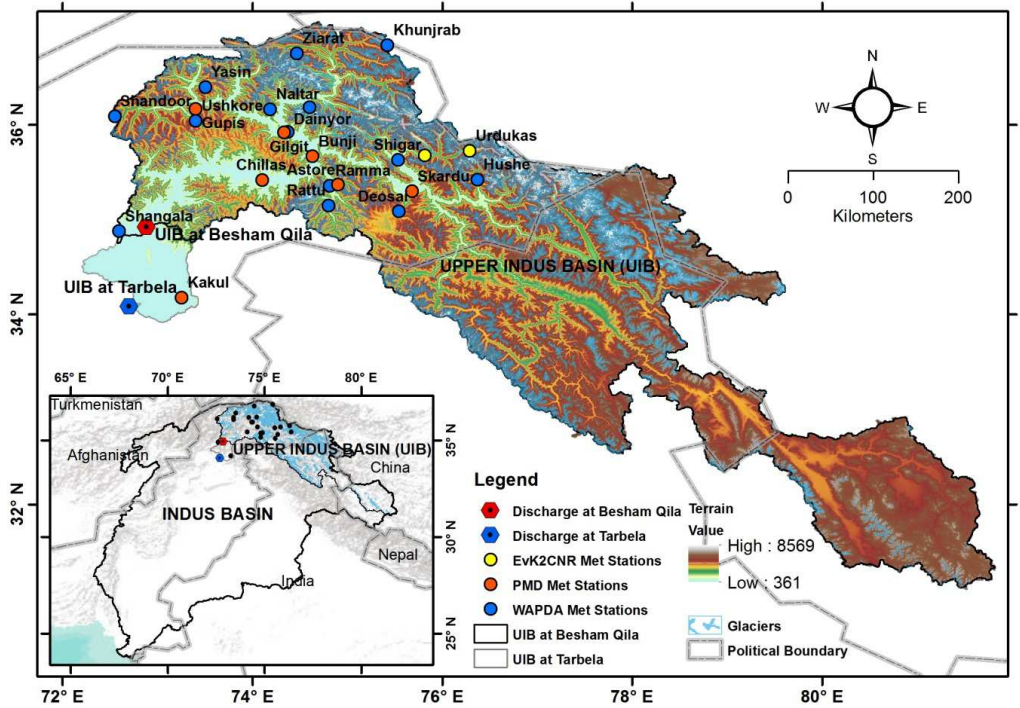

Figure 1: Study Area, Upper Indus Basin (UIB) and meteorological station networks.

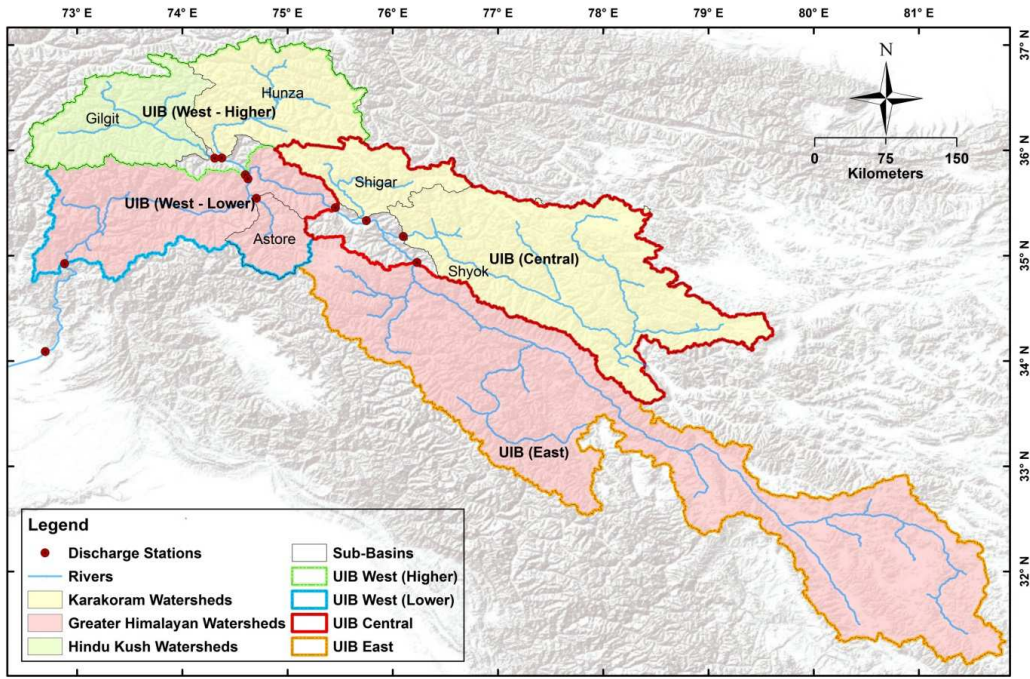

Figure 2: Gauged basins and gauging stations along with considered regions for field significance.

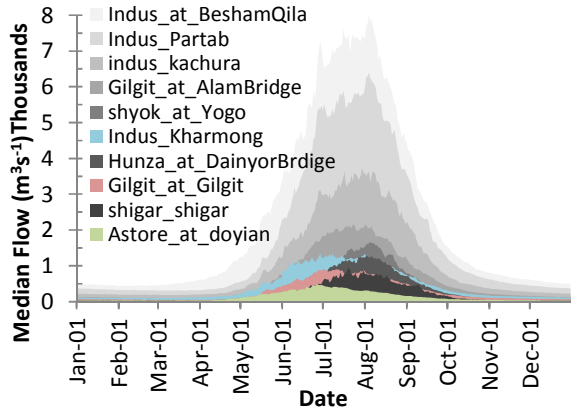

Figure 3: Long-term median hydrograph for ten key gauging stations separating the sub-basins of UIB having either mainly snow-fed (shown in color) or mainly glacier-fed hydrological regimes (shown in grey shades).

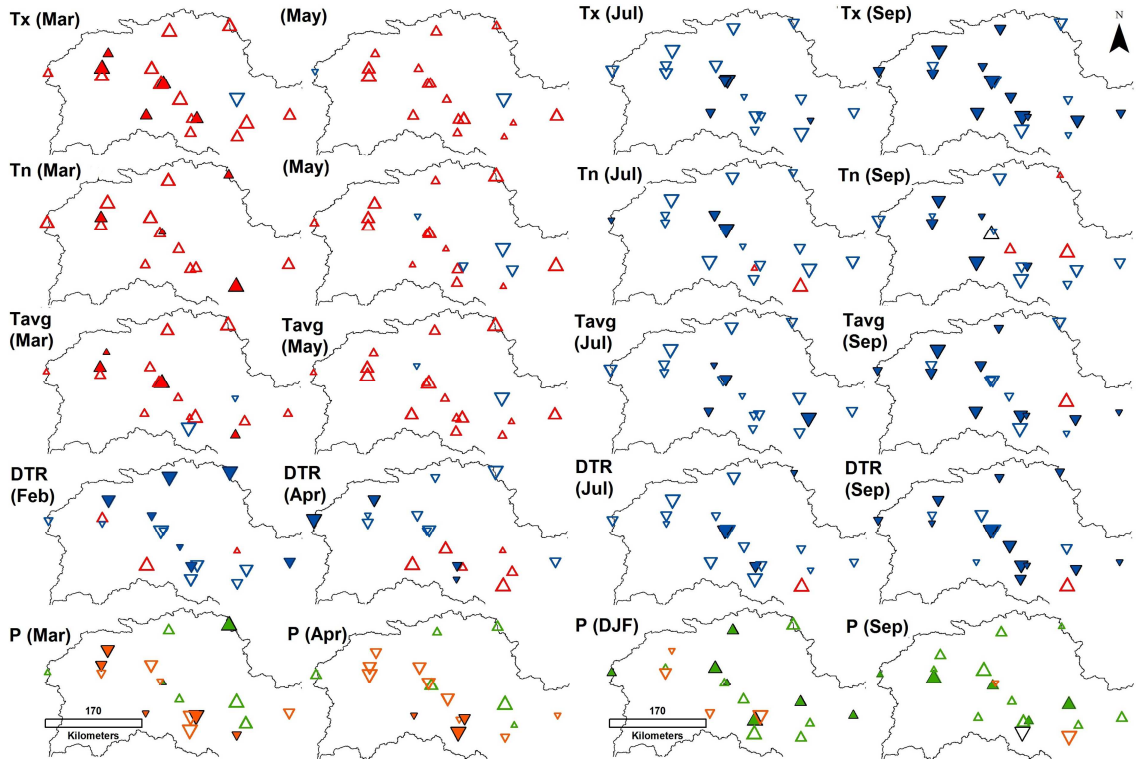

Figure 4: Trend per time step of cooling (downward) and warming (upward) in Tx, Tn and Tavg, and increase (upward) and decrease (downward) in DTR and in P for select months and seasons. Statistically significant trends at ≥ 90% level are shown in solid triangle, the rest in hollow triangles.

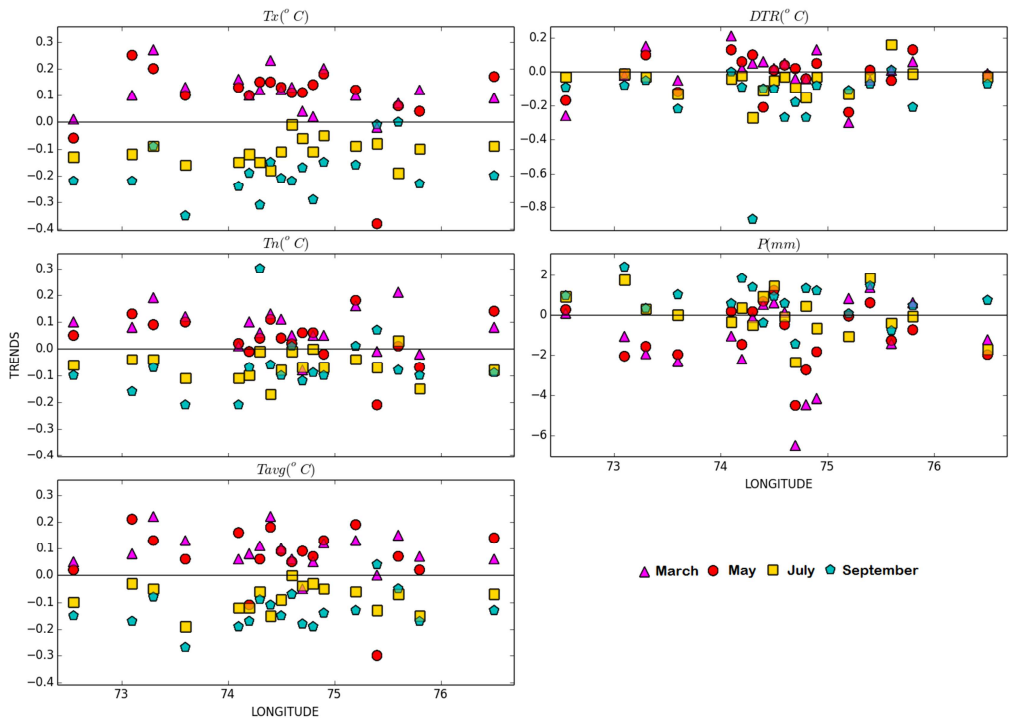

Figure 5: Hydroclimatic trends per unit time for the period 1995-2012 against longitude.

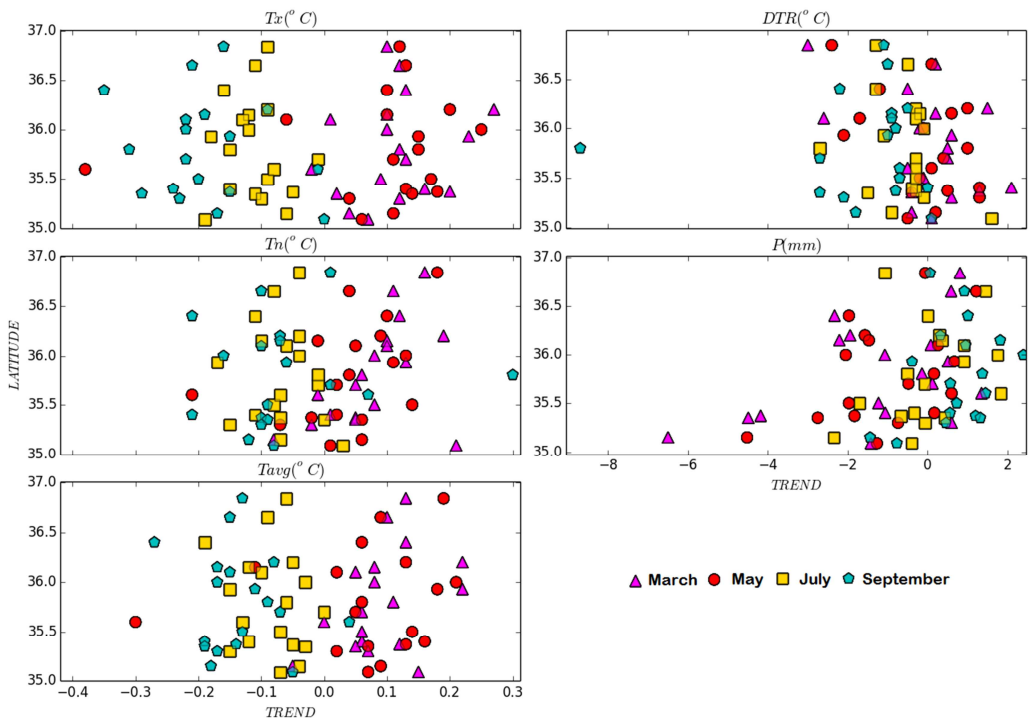

Figure 6: Hydroclimatic trends per unit time for the period 1995-2012 against latitude. Here for DTR only overall trend changes over the whole 1995-2012 period are shown.

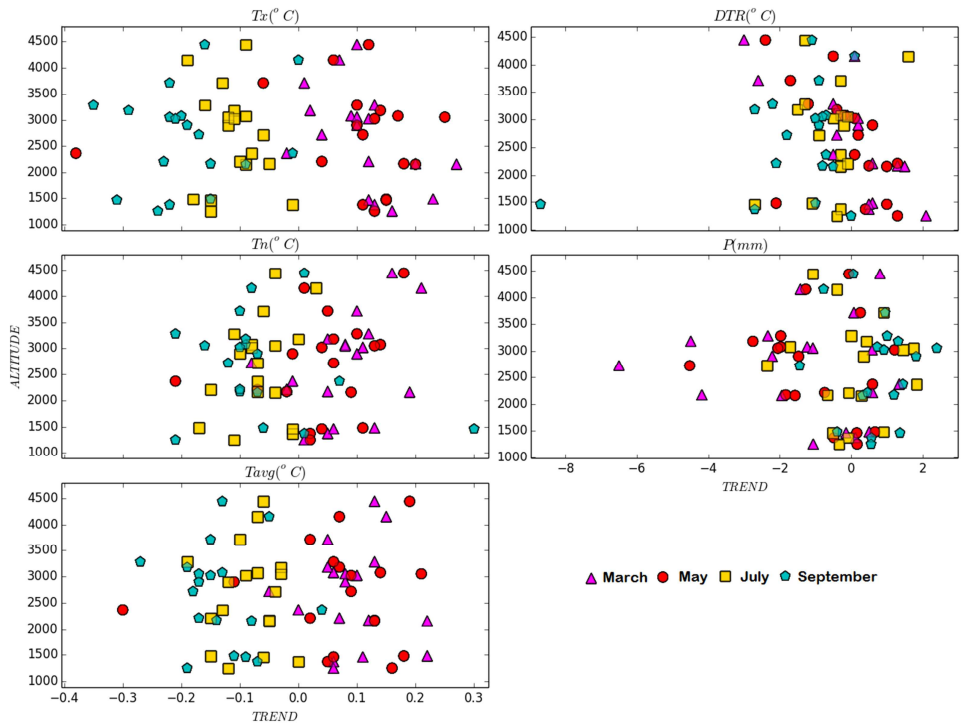

Figure 7: Same as Figure 6 but against altitude.